\documentclass[12pt]{article}

\usepackage{axodraw}
\usepackage{amssymb}
\usepackage{epsfig}

\catcode`@=11
\def\citer{\@ifnextchar [{\@tempswatrue\@citexr}{\@tempswafalse\@citexr[]}}
 
\def\@citexr[#1]#2{\if@filesw\immediate\write\@auxout{\string\citation{#2}}\fi
  \def\@citea{}\@cite{\@for\@citeb:=#2\do
    {\@citea\def\@citea{--\penalty\@m}\@ifundefined
       {b@\@citeb}{{\bf ?}\@warning
       {Citation `\@citeb' on page \thepage \space undefined}}%
\hbox{\csname b@\@citeb\endcsname}}}{#1}}
\catcode`@=12
 
\reversemarginpar

\newcommand{\tgb}{{\rm tg} \beta}

\newcommand{\sgl}{{\tilde{g}}}
\newcommand{\sbottom}{{\tilde{b}}}
\newcommand{\sq}{{\tilde{q}}}
\newcommand{\sqb}{{\overline{\tilde{q}}}}

\begin{document}

\renewcommand{\thefootnote}{\fnsymbol{footnote} }

\vskip-1.0cm

\begin{flushright}
PSI--PR--11--01
\end{flushright}

\begin{center}
{\large\sc Neutral Higgs Boson Decays to Squark Pairs reanalyzed}
\end{center}

\begin{center}
E.~Accomando$^1$, G.~Chachamis$^2$, F.~Fugel$^2$, M.~Spira$^2$ and
M.~Walser$^{2,3}$
\end{center}

\begin{center}
{\it \small
$^1$ University of Southampton, Theory Group, Southampton SO17 1BJ,
United Kingdom \\
$^2$ Paul Scherrer Institut, CH-5232 Villigen PSI, Switzerland \\
$^3$ Institute for Theoretical Physics, ETH Z\"urich, CH--8093 Z\"urich,
Switzerland}
\end{center}

\begin{abstract}
We analyze neutral Higgs boson decays into squark pairs in the minimal
supersymmetric extension of the Standard Model and improve previous
analyses. In particular the treatment of potentially large higher-order
corrections originating from the soft SUSY breaking parameters $A_b$, the
trilinear Higgs coupling to sbottoms, and $\mu$, the Higgsino mass
parameter, is investigated. The remaining theoretical uncertainties
including the SUSY-QCD corrections are analyzed quantitatively.
\end{abstract}

\def\thefootnote{\arabic{footnote}}
\setcounter{footnote}{0}

\section{Introduction}
The search for Higgs bosons plays one of the most important roles at
high-energy collider experiments at the Tevatron and the LHC. The Higgs
boson is the remnant of electroweak symmetry breaking in the scalar
Higgs sector of the Standard Model (SM) and its supersymmetric
extensions. The minimal supersymmetric extension of the SM (MSSM)
requires the introduction of two Higgs doublets in order to preserve
supersymmetry \cite{twoiso}. This leads to the existence of five
elementary Higgs particles, two CP-even ($h,H$), one CP-odd ($A$) and
two charged ($H^\pm$) states. At lowest order (LO) all couplings and masses
of the MSSM Higgs sector are described by two independent input
parameters, which are usually chosen as $\tgb=v_2/v_1$, the ratio of the
two vacuum expectation values $v_{1,2}$, and the pseudoscalar
Higgs-boson mass $M_A$. At LO, the light scalar Higgs mass $M_h$ has to
be smaller than the $Z$-boson mass $M_Z$. Including the one-loop and
dominant two-loop corrections the upper bound is increased to
$M_h\lesssim 135$ GeV \cite{mssmrad}. Recent first three-loop
results confirm this upper bound within less than 1 GeV \cite{mssmrad3}.
The couplings of the various neutral Higgs bosons to fermions and gauge
bosons depend on the mixing angles $\alpha$ and $\beta$.  Normalized to
the SM Higgs couplings, they are listed in Table~\ref{tb:hcoup}.
\begin{table}[hbt]
\renewcommand{\arraystretch}{1.5}
\begin{center}
\begin{tabular}{|lc||ccc|} \hline
\multicolumn{2}{|c||}{$\Phi$} & $g^\Phi_u$ & $g^\Phi_d$ &  $g^\Phi_V$ \\
\hline \hline
SM~ & $H$ & 1 & 1 & 1 \\ \hline
MSSM~ & $h$ & $\cos\alpha/\sin\beta$ & $-\sin\alpha/\cos\beta$ &
$\sin(\beta-\alpha)$ \\
& $H$ & $\sin\alpha/\sin\beta$ & $\cos\alpha/\cos\beta$ &
$\cos(\beta-\alpha)$ \\
& $A$ & $ 1/\tgb$ & $\tgb$ & 0 \\ \hline
\end{tabular}
\renewcommand{\arraystretch}{1.2}
\caption[]{\label{tb:hcoup} \it Higgs couplings in the MSSM to fermions
and gauge bosons [$V=W,Z$] relative to SM couplings. The subscripts
$u,d$ denote up- and down-type fermions.}
\end{center}
\end{table}
For large values of $\tgb$ the down-type Yukawa couplings are strongly
enhanced, while the up-type Yukawa couplings are suppressed, unless the
light (heavy) scalar Higgs mass ranges at its upper (lower) bound, where
the couplings become SM-like.

The negative direct searches for the MSSM Higgs bosons at LEP2 yield
lower bounds of $M_{h,H} > 92.8$ GeV and $M_A > 93.4$ GeV. For a SUSY
scale $M_{SUSY}=1$ TeV the range $0.7 < \tgb < 2.0$ in the MSSM is
excluded by the Higgs searches at the LEP2 experiments \cite{lep2}.
Presently and in the future, Higgs bosons can be searched for at the
Fermilab Tevatron \cite{Carena:2000yx}, a proton-antiproton collider
with a center-of-mass energy of 1.96 TeV, and the proton-proton
Large Hadron Collider (LHC) with up to 14 TeV center-of-mass energy
\cite{atlas_cms_tdrs}.

The MSSM Higgs bosons couple to squarks, too. If the Higgs masses are
large enough to allow for decays into third-generation squark-antisquark
pairs, these decay modes acquire sizable branching ratios in many MSSM
scenarios \cite{ohmann}. Thus, for a reliable theoretical treatment of
these supersymmetric Higgs decay modes, higher order corrections have to
be computed and included appropriately. In the past the full SUSY--QCD
corrections \cite{h2sqcorr} and the full SUSY--electroweak corrections
\cite{h2sqelw} to the Higgs decays into squarks have been calculated.
In particular regions of the MSSM parameter space the corrections turned
out to be so large that a reliable prediction was not possible without
further refinements. A first attempt to solve this problem has been
undertaken in Ref.~\cite{hsqimp} by starting from a more consistent
treatment of the squark masses and couplings at next-to-leading order
(NLO).  However, the approach of Ref.~\cite{hsqimp} does not provide a
treatment of the squark mixing starting from the running soft
SUSY-breaking parameters of the squark sector only. The topic of this
work is a complete and consistent determination of the MSSM squark
sector and the Higgs couplings to squarks from running $\overline{MS}$
input parameters in the context of SUSY--QCD corrections\footnote{We
have chosen the $\overline{MS}$ scheme for convenience while our
analysis could easily be translated to the $\overline{DR}$ scheme as
used in spectrum generators. The $\overline{MS}$ parameters can be
obtained from the $\overline{DR}$ ones by simple relations
\cite{susyrest} and {\it vice versa}.}. The input parameters can be
obtained from the renormalization group equations in the framework of a
certain SUSY-breaking mechanism.  Our treatment can also be used for a
consistent global fit to supersymmetric observables which include
observables based on the stop and sbottom sectors of the MSSM.

This paper is organized as follows. Section \ref{sc:masscoup} describes
the systematic determination of the squark sector at NLO, while in
Section \ref{sc:h2sqsq} we summarize the SUSY--QCD corrections to the
MSSM Higgs decays into squarks. Numerical results are presented in
Section \ref{sc:numerics}. Finally in Section \ref{sc:conclusions} we
conclude.

\section{Squark masses and couplings} \label{sc:masscoup}
In this section, we describe in detail the determination of the
stop and sbottom masses and their couplings to the MSSM Higgs bosons at
LO, and extend the setup to NLO consistently. The NLO expressions will be
derived for soft supersymmetry breaking parameters given in the
$\overline{MS}$ scheme.

\subsection{Sfermion masses and couplings at LO}
The scalar partners $\tilde f_{L,R}$ of the left- and right-handed fermion
components mix with each other. The mass eigenstates $\tilde f_{1,2}$ of the
sfermions $\tilde f$ are related to the current eigenstates $\tilde f_{L,R}$
by mixing angles $\theta_f$,
\begin{eqnarray}
\tilde f_1 & = & \tilde f_L \cos\theta_f + \tilde f_R \sin \theta_f \nonumber \\
\tilde f_2 & = & -\tilde f_L\sin\theta_f + \tilde f_R \cos \theta_f \, ,
\label{eq:sfmix}
\end{eqnarray}
which are proportional to the masses of the ordinary fermions, see
Eq.(\ref{eq:sqmix0}). Thus mixing effects are only important for the
third-generation sfermions $\tilde t, \tilde b, \tilde \tau$, the mass
matrix of which is given by
\begin{equation}
{\cal M}_{\tilde f} = \left[ \begin{array}{cc}
\tilde M_{\tilde f_L}^2 + m_f^2 & m_f (A_f-\mu r_f) \\
m_f (A_f-\mu r_f) & \tilde M_{\tilde f_R}^2 + m_f^2 \end{array} \right] \, ,
\end{equation}
with the parameters $r_b = r_\tau = 1/r_t = \tgb$. The parameter $A_f$
denotes the trilinear sfermion coupling of the soft supersymmetry
breaking part of the Lagrangian, while $\mu$ is the Higgsino mass
parameter and $m_f$ the corresponding fermion mass. The $D$-terms have
been absorbed in the parameters $\tilde M_{\tilde f_{L/R}}$,
\begin{eqnarray}
\tilde M^2_{\tilde f_{L/R}} & = & M^2_{\tilde f_{L/R}} + D_{\tilde f_{L/R}}
\nonumber \\
D_{\tilde f_L} & = & M_Z^2 (I^f_{3L} - e_f \sin^2\theta_W) \cos 2\beta
\nonumber \\
D_{\tilde f_R} & = & M_Z^2 e_f \sin^2\theta_W \cos 2\beta \, ,
\label{eq:dterms}
\end{eqnarray}
where $M_{\tilde f_{L/R}}$ denotes the sfermion masses of the soft
supersymmetry breaking part of the Lagrangian.  Consequently the mixing
angles acquire the form
\begin{equation}
\sin 2\theta_f = \frac{2m_f (A_f-\mu r_f)}{m_{\tilde f_1}^2 - m_{\tilde f_2}^2}
~~~,~~~
\cos 2\theta_f = \frac{\tilde M_{\tilde f_L}^2 - \tilde M_{\tilde f_R}^2}
{m_{\tilde f_1}^2 - m_{\tilde f_2}^2}
\label{eq:sqmix0}
\end{equation}
and the masses of the squark eigenstates are given by
\begin{equation}
m_{\tilde f_{1,2}}^2 = m_f^2 + \frac{1}{2}\left[ \tilde M_{\tilde f_L}^2 +
\tilde M_{\tilde f_R}^2 \mp \sqrt{(\tilde M_{\tilde f_L}^2 -
\tilde M_{\tilde f_R}^2)^2 + 4m_f^2 (A_f - \mu r_f)^2} \right] \, .
\end{equation}
In the current eigenstate basis the neutral Higgs couplings to sfermions
read as\footnote{In our notation the first index of the neutral Higgs
couplings to sfermions defines the incoming and the second index the
outgoing sfermion at the corresponding vertex.}
\begin{eqnarray}
g_{\tilde f_L \tilde f_L}^\Phi & = & m_f^2 g_1^\Phi + M_Z^2 (I_{3f}
- e_f\sin^2\theta_W) g_2^\Phi \nonumber \\
g_{\tilde f_R \tilde f_R}^\Phi & = & m_f^2 g_1^\Phi + M_Z^2 e_f\sin^2\theta_W
g_2^\Phi \nonumber \\
g_{\tilde f_L \tilde f_R}^\Phi & = & -\frac{m_f}{2} (\mu g_3^\Phi
- A_f g_4^\Phi)
\label{eq:hsfcouprl}
\end{eqnarray}
with the couplings $g_i^\Phi~(i=1,\ldots,4)$ listed in Table
\ref{tb:hsfcoup}. For the scalar Higgs bosons $h,H$ the couplings to
sfermions are symmetric, i.e.  $g_{\tilde f_R \tilde f_L}^{h,H} =
g_{\tilde f_L \tilde f_R}^{h,H}$. For the pseudoscalar Higgs boson $A$
the diagonal couplings $g_{\tilde f_L \tilde f_L}^A$ and $g_{\tilde f_R
\tilde f_R}^A$ vanish, while the off-diagonal couplings are
antisymmetric, i.e.~$g_{\tilde f_R \tilde f_L}^A=-g_{\tilde f_L \tilde
f_R}^A$. The corresponding couplings to the sfermion mass eigenstates
$\tilde f_{1,2}$ are given by
\begin{eqnarray}
g_{\tilde f_1 \tilde f_1}^{h,H} & = & g_{\tilde f_L \tilde f_L}^{h,H}
\cos^2\theta_f + g_{\tilde f_R \tilde f_R}^{h,H} \sin^2\theta_f +
g_{\tilde f_L \tilde f_R}^{h,H} \sin 2\theta_f \nonumber \\
g_{\tilde f_2 \tilde f_2}^{h,H} & = & g_{\tilde f_L \tilde f_L}^{h,H}
\sin^2\theta_f + g_{\tilde f_R \tilde f_R}^{h,H} \cos^2\theta_f -
g_{\tilde f_L \tilde f_R}^{h,H} \sin 2\theta_f \nonumber \\
g_{\tilde f_1 \tilde f_2}^{h,H} & = & g_{\tilde f_2 \tilde f_1}^{h,H}
= \frac{1}{2}(g_{\tilde f_R \tilde f_R}^{h,H}
- g_{\tilde f_L \tilde f_L}^{h,H}) \sin 2\theta_f + g_{\tilde f_L \tilde
  f_R}^{h,H} \cos 2\theta_f \nonumber \\
g_{\tilde f_1 \tilde f_1}^A & = & g_{\tilde f_2 \tilde f_2}^A = 0
\nonumber \\
g_{\tilde f_1 \tilde f_2}^A & = & -g_{\tilde f_2 \tilde f_1}^A =
g_{\tilde f_L \tilde f_R}^A \, .
\label{eq:couprot}
\end{eqnarray}
For a consistent NLO calculation the relations for the sfermion masses,
mixing angles and couplings have to be extended to NLO consistently. In
the following we will concentrate on the stop and sbottom sectors at NLO
SUSY--QCD.
\begin{table}[hbt]
\renewcommand{\arraystretch}{1.5}
\begin{center}
\begin{tabular}{|l|c||c|c|c|c|} \hline
$\tilde f$ & $\Phi$ & $g^\Phi_1$ & $g^\Phi_2$ & $g^\Phi_3$ & $g^\Phi_4$ \\
\hline \hline
& $h$ & $\cos\alpha/\sin\beta$ & $-\sin(\alpha+\beta)$ &
$-\sin\alpha/\sin\beta$ & $\cos\alpha/\sin\beta$ \\
$\tilde u$ & $H$ & $\sin\alpha/\sin\beta$ & $\cos(\alpha+\beta)$ &
$\cos\alpha/\sin\beta$ & $\sin\alpha/\sin\beta$ \\
& $A$ & 0 & 0 & 1 & $-1/\tgb$ \\ \hline
& $h$ & $-\sin\alpha/\cos\beta$ & $-\sin(\alpha+\beta)$ &
$\cos\alpha/\cos\beta$ & $-\sin\alpha/\cos\beta$ \\
$\tilde d$ & $H$ & $\cos\alpha/\cos\beta$ & $\cos(\alpha+\beta)$ &
$\sin\alpha/\cos\beta$ & $\cos\alpha/\cos\beta$ \\
& $A$ & 0 & 0 & 1 & $-\tgb$ \\ \hline
\end{tabular} 
\renewcommand{\arraystretch}{1.2}
\caption[]{\label{tb:hsfcoup}
\it Coefficients of the neutral MSSM Higgs couplings to sfermion pairs.
The symbols $\tilde u,\tilde d$ denote up- and down-type sfermions.}
\end{center}
\end{table}

\subsection{Stops and Sbottoms at NLO}
The soft supersymmetry-breaking parameters $\overline{M}_{\tilde
q_{L,R}}(Q_0)$ and $\bar A_q(Q_0)$ will be introduced as $\overline{MS}$
parameters at an input scale $Q_0$ which will in general be of the order
of the SUSY scale.

In order to resum large corrections for large values of $\tgb$ we define
the sbottom masses and couplings to the Higgs bosons in terms of the
effective bottom mass\footnote{Note that this definition of the
effective bottom quark mass differs from the $\overline{DR}$ definition
which has been used in Ref.~\cite{hsqimp}. Our effective bottom mass
runs with five active flavours consistently, i.e.~all heavier particles
are decoupled, while the $\overline{DR}$ definition requires running
with the contributions of all supersymmetric particles. The scale of the
strong coupling $\alpha_s$ in $\Delta_b$ within the effective bottom
mass is identified with the average mass of the sbottom and gluino
states in order to account for the NNLO corrections to a large extent
\cite{noth} and is thus different from the scale $Q$.}
\cite{hsqimp,deltamb,guasch}
\begin{eqnarray}
\hat m_b(Q) & = & \frac{\overline{m}_b(Q)}{1+\Delta_b} \nonumber \\
\Delta_b & = & \frac{C_F}{2}~\frac{\alpha_s}{\pi}~M_{\sgl}~\mu~\tgb~
I(m^2_{\sbottom_1},m^2_{\sbottom_2},M^2_{\sgl}) \nonumber \\
I(a,b,c) & = & \frac{\displaystyle ab\log\frac{a}{b} + bc\log\frac{b}{c}
+ ca\log\frac{c}{a}}{(a-b)(b-c)(a-c)}
\label{eq:deltab}
\end{eqnarray}
with $C_F=4/3$, where $\overline{m}_b(Q)$ denotes the running bottom
mass in the $\overline{MS}$ scheme at the scale $Q$ and
$m_{\sbottom_1},m_{\sbottom_2},M_{\sgl}$ are the sbottom and gluino pole
masses respectively.  The effective top mass is identified with the
running $\overline{MS}$ top mass
\begin{equation}
\hat m_t(Q) = \overline{m}_t(Q) \, .
\label{eq:tmass}
\end{equation}
The use of these effective masses corresponds to the consistent
inclusion of the resummed and RG-improved Yukawa couplings of the Higgs
decays $\phi^0\to t\bar t, b\bar b$ everywhere in the squark sectors,
too\footnote{A similar approach has meanwhile been pursued in
Ref.~\cite{crivellin} for the Higgsino couplings to fermions and
sfermions, too.}. The stop/sbottom mass matrix at LO is then given by
($q=t,b$)
\begin{equation}
{\cal M}_{\tilde q} = \left[ \begin{array}{cc}
\tilde{\overline{M}}_{\tilde q_L}^2(Q_0) + \hat m_q^2(Q_0) & \hat
m_q(Q_0)
[\bar{A}_q(Q_0)-\mu r_q] \\ \hat m_q(Q_0) [\bar{A}_q(Q_0)-\mu r_q]
& \tilde{\overline{M}}_{\tilde q_R}^2(Q_0) + \hat m_q^2(Q_0)
\end{array} \right] \, ,
\end{equation}
where $\bar{A}_q(Q_0)$ denotes the running trilinear $\overline{MS}$
coupling at the scale $Q_0$ and $r_b=1/r_t=\tgb$. Analogous to
Eqs.~(\ref{eq:dterms}) the $D$-terms $D_{\tilde q_{L/R}}$ have been
absorbed in the parameters $\tilde{\overline{M}}_{\tilde q_{L/R}}(Q_0)$,
\begin{equation}
\tilde{\overline{M}}^2_{\tilde q_{L/R}}(Q_0) = \overline{M}^2_{\tilde
q_{L/R}}(Q_0) + D_{\tilde q_{L/R}} \, .
\label{eq:dterms_b}
\end{equation}
The stop/sbottom mass matrix is modified by higher-order corrections in
the diagonal and off-diagonal entries. While the corrections to the
off-diagonal entries will be treated by the renormalization of the
mixing angles, we compensate the radiative corrections to the diagonal
matrix elements by shifts in the soft mass parameters
$\overline{M}_{\tilde q_{L/R}}(Q_0)$,
\begin{equation}
M_{\tilde q_{L/R}}^2(Q_0) = \overline{M}^2_{\tilde q_{L/R}}(Q_0)
+ \Delta \overline{M}_{\tilde q_{L/R}}^2 \qquad , \qquad 
\tilde M_{\tilde q_{L/R}}^2(Q_0) = \tilde{\overline{M}}^2_{\tilde q_{L/R}}(Q_0)
+ \Delta \overline{M}_{\tilde q_{L/R}}^2
\label{eq:shift}
\end{equation}
in order to arrive at simple expressions at NLO for the stop/sbottom
masses.  Using the squark pole masses $m_{\tilde q_{1/2}}$ the
tree-level definition $\tilde \theta_q$ of the mixing angle is given by
\begin{equation}
\sin 2\tilde\theta_q = \frac{2\hat m_q(Q_0) [\bar{A}_q(Q_0)-\mu r_q]}{ m_{\tilde
q_1}^2 - m_{\tilde q_2}^2} \quad , \quad
\cos 2\tilde\theta_q = \frac{\tilde M_{\tilde q_L}^2(Q_0) - \tilde M_{\tilde
q_R}^2(Q_0)} {m_{\tilde q_1}^2 - m_{\tilde q_2}^2} \, ,
\label{eq:sqmix}
\end{equation}
where the shifted squark mass parameters $\tilde M_{\tilde
q_{L/R}}(Q_0)$ have been used with the radiatively corrected squark pole
masses.  This definition will be used as the tree-level-like mixing
angle with loop-corrected squark masses $m_{\tilde q_{1/2}}$ at NLO,
too, but will only play the role of an auxiliary parameter in our
analysis as will be explained later.

At NLO the masses of the stop/sbottom eigenstates are given by
\begin{eqnarray}
m_{\tilde q_{1/2}}^2 & = & \hat m_q^2(Q_0) + \frac{1}{2}\left[
\tilde{\overline{M}}_{\tilde q_L}^2(Q_0) + \tilde{\overline{M}}_{\tilde
q_R}^2(Q_0) \right. \nonumber \\
& & \left. \hspace*{1.5cm} \mp \sqrt{[\tilde{\overline{M}}_{\tilde
q_L}^2(Q_0) - \tilde{\overline{M}}_{\tilde q_R}^2(Q_0)]^2 + 4\hat
m_q^2(Q_0) [\bar{A}_q(Q_0) - \mu r_q]^2} \right] + \Delta m_{\tilde
q_{1/2}}^2 \nonumber \\
\Delta m_{\tilde q_{1/2}}^2 & = & \Sigma_{11/22}(m_{\tilde q_{1/2}}^2) +
\delta \hat m_{\tilde q_{1/2}}^2
\label{eq:sqmass}
\end{eqnarray}
The diagonal parts $\Sigma_{11/22}$ of the stop/sbottom self-energies
can be calculated from the diagrams in Fig.~\ref{fg:sqself},
\begin{figure}[htb]
\begin{center}
\SetScale{0.8}
\begin{picture}(130,40)(130,-10)
\DashLine(0,0)(100,0){5}
\GOval(50,0)(10,20)(0){0.5}
\DashLine(130,0)(230,0){5}
\GlueArc(180,0)(20,0,180){4}{5}
\DashLine(260,0)(360,0){5}
\ArrowLine(290,0)(330,0)
\CArc(310,0)(20,0,180)
\GlueArc(310,0)(20,0,180){4}{5}
\DashLine(390,0)(490,0){5}
\DashCArc(440,20)(20,0,360){5}
\Vertex(440,0){2}
\put(88,-3){$=$}
\put(193,-3){$+$}
\put(297,-3){$+$}
\put(10,6){$\tilde q_i$}
\put(70,6){$\tilde q_j$}
\put(113,6){$\tilde q_i$}
\put(143,-13){$\tilde q$}
\put(143,25){$g$}
\put(175,6){$\tilde q_j$}
\put(218,6){$\tilde q_i$}
\put(248,-12){$q$}
\put(248,25){$\tilde g$}
\put(280,6){$\tilde q_j$}
\put(323,6){$\tilde q_i$}
\put(351,35){$\tilde q$}
\put(383,6){$\tilde q_j$}
\end{picture} \\
\caption{\label{fg:sqself} \it One-loop contributions to the squark
self-energies.}
\vspace*{-0.5cm}
\end{center}
\end{figure}
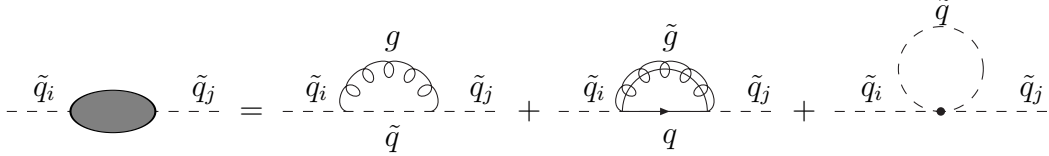
\begin{eqnarray}
\Sigma_{11/22}(m_{\tilde q_{1/2}}^2) & = & C_F\frac{\alpha_s}{\pi}
\frac{1}{4} \left\{
-(1+\cos^2 2\tilde\theta_q) A_0(m_{\tilde q_{1/2}}) - \sin^2 2\tilde\theta_q
A_0(m_{\tilde q_{2/1}}) \right.  \nonumber \\
& & + 2A_0(M_\sgl) + 2A_0[\hat m_q(Q_0)] + 4 m_{\tilde q_{1/2}}^2
B_0(m_{\tilde q_{1/2}}^2;0,m_{\tilde q_{1/2}}) \label{eq:sigii} \\
& + & \left. \!\!\!\!\! 2\!\left[ M_\sgl^2 + \hat m^2_q(Q_0) - m_{\tilde
q_{1/2}}^2 \!\! \mp 2M_\sgl \hat m_q(Q_0) \sin 2\tilde\theta_q \right] \!
B_0[m_{\tilde q_{1/2}}^2;M_\sgl,\hat m_q(Q_0)] \right\} \, . \nonumber
\end{eqnarray}
The one-loop integrals are defined as \cite{passvelt}
\begin{eqnarray}
A_0(m) & = & \int \frac{d^n k}{(2\pi)^n} \frac{-i(4\pi)^2
\bar\mu^{2\epsilon}}{k^2-m^2} \nonumber \\
B_0(p^2;m_1,m_2) & = & \int \frac{d^n k}{(2\pi)^n} \frac{-i(4\pi)^2
\bar\mu^{2\epsilon}}{[k^2-m_1^2][(k+p)^2-m_2^2]} \nonumber \\
B_1(p^2;m_1,m_2) & = & \frac{1}{2p^2}\left\{ A_0(m_1)-A_0(m_2) -
(p^2+m_1^2-m_2^2) B_0(p^2;m_1,m_2) \right\} \, .
\end{eqnarray}
The parameter $\bar \mu$ denotes the 't Hooft mass of dimensional
regularization. The counter terms $\delta \hat m_{\tilde q_{1,2}}^2$ are
given by
\begin{eqnarray}
\delta \hat m_{\tilde q_{1/2}}^2 & = & 2\hat m_q(Q_0)\delta \hat m_q +
\frac{1}{2}\left\{ \delta \overline{M}_{\tilde q_L}^2 + \delta
\overline{M}_{\tilde q_R}^2 \pm \left[ (\delta \overline{M}_{\tilde
q_L}^2-\delta \overline{M}_{\tilde q_R}^2) \cos 2\tilde \theta_q \right.
\right. \nonumber \\
& & \left. \left. \hspace*{1.5cm} + \left( \frac{\delta \hat m_q}{\hat
m_q(Q_0)} + \frac{\delta \bar A_q}{\bar A_q(Q_0)-\mu r_q} \right)
(m_{\tilde q_1}^2 - m_{\tilde q_2}^2) \sin^2 2\tilde \theta_q \right]
\right\} \nonumber \\
& = & - C_F\frac{\alpha_s}{\pi} \Gamma(1+\epsilon) (4\pi)^\epsilon
\left\{ \frac{1}{\epsilon} + \log\frac{\bar \mu^2}{Q_0^2} \right\}
\left\{M^2_\sgl \mp M_\sgl \hat m_q(Q_0) \sin 2\tilde \theta_q) \right\}
\nonumber \\
& & + \frac{\delta \hat m_q}{\hat m_q(Q_0)} \left\{ 2\hat m^2_q(Q_0) \mp
\frac{1}{2}(m_{\tilde q_2}^2 - m_{\tilde q_1}^2) \sin^2 2\tilde
\theta_q \right\} \, ,
\label{eq:dmsq}
\end{eqnarray}
where the tree-level-like mixing angle $\tilde \theta_q$ of
Eq.~(\ref{eq:sqmix}) has been used and the parameters
$\overline{M}_{\sq_{L/R}}^2$ and $\bar A_q$ have been renormalized in
the $\overline{MS}$ scheme,
\begin{eqnarray}
\delta \overline{M}^2_{\sq_{L/R}} & = & -C_F \frac{\alpha_s}{\pi}
\Gamma(1+\epsilon) (4\pi)^\epsilon M^2_\sgl \left\{ \frac{1}{\epsilon} +
\log \frac{\bar \mu^2}{Q_0^2} \right\} \nonumber \\
\delta \bar A_q & = & C_F \frac{\alpha_s}{\pi} \Gamma(1+\epsilon)
(4\pi)^\epsilon M_\sgl \left\{ \frac{1}{\epsilon} + \log \frac{\bar
\mu^2}{Q_0^2} \right\} \, .
\label{eq:daq}
\end{eqnarray}
The counter term of the effective quark mass $\hat m_q(Q)$ for $Q=Q_0$
is given by\footnote{The quark mass counter term is different from the
$\overline{\rm DR}$ expression used in Ref.~\cite{hsqimp}.}
\begin{eqnarray}
\frac{\delta \hat m_q}{\hat m_q(Q)} & = & -C_F\frac{\alpha_s}{\pi}
\Gamma(1+\epsilon) (4\pi)^\epsilon \frac{3}{4} \left\{\frac{1}{\epsilon}
+\log \frac{\bar \mu^2}{Q^2} + \delta_{SUSY} \right\} + \Delta_q
\nonumber \\
& - & C_F \frac{\alpha_s}{4\pi} \left\{ B_1[\hat
m_q^2(Q);M_\sgl,m_{\tilde q_1}] + B_1[\hat m_q^2(Q);M_\sgl,m_{\tilde
q_2}] \right. \nonumber \\
& & \left. \hspace*{0.7cm} +2 M_\sgl (\bar A_q-\mu r_q) \frac{B_0[\hat
m_q^2(Q);M_\sgl,m_{\tilde q_1}] - B_0[\hat m_q^2(Q);M_\sgl,m_{\tilde
q_2}]}{m^2_{\tilde q_1}-m^2_{\tilde q_2}}\right\} \, ,
\label{eq:dmq}
\end{eqnarray}
where $\delta_{SUSY} = 1/3$ is a finite counter term required to restore
the supersymmetric relations between the Higgs boson couplings to
top/bottom quarks and stops/sbottoms within dimensional regularization
\cite{susyrest}.  The term $\Delta_q$ denotes the correction $\Delta_b$
of Eq.~(\ref{eq:deltab}) for the bottom mass, while it vanishes for the
top mass case, i.e.~$\Delta_t=0$. The term $\Delta_b$ in the counter
term of the bottom mass cancels the leading term for large values of
$\tgb$ in the last line of Eq.~(\ref{eq:dmq}) so that this counter term
is free of large corrections. Using the NLO corrected squark pole masses
of Eq.~(\ref{eq:sqmass}) and the tree-level-like mixing angle
$\tilde\theta_q$ of Eq.~(\ref{eq:sqmix}), the shifted squared soft
SUSY-breaking squark mass parameters $\tilde M^2_{\tilde q_{L/R}}(Q_0) =
\tilde{\overline{M}}^2_{\tilde q_{L/R}}(Q_0) + \Delta
\overline{M}^2_{\tilde q_{L/R}}$ can be obtained from the
relations\footnote{These equations differ from the corresponding
inconsistent relations in terms of the on-shell definition of the mixing
angle used in \cite{hsqimp}. Our approach starts from the consistent
tree-level relations and incorporates the anti-Hermitian definition of
the mixing angle in a finite shift from the tree-level value
$\tilde\theta_q$. An alternative and equivalent option would be to
compensate the shift $\Delta \tilde\theta_q$ by a shifted trilinear
coupling $A_q$. Moreover, it should be noted that the shifted parameters
$M_{\tilde t_L}(Q_0)$ and $M_{\tilde b_L}(Q_0)$ are not equal to each
other any more in accordance with the different radiative corrections to
the stop and sbottom masses.},
\begin{eqnarray}
\tilde M^2_{\tilde q_L}(Q_0) & = & M_{\tilde q_L}^2(Q_0) + D_{\tilde q_L} =
m^2_{\tilde q_1} \cos^2 \tilde\theta_q + m^2_{\tilde q_2} \sin^2
\tilde\theta_q - \hat m_q^2(Q_0) \nonumber \\
\tilde M^2_{\tilde q_R}(Q_0) & = & M_{\tilde q_R}^2(Q_0) + D_{\tilde q_R} =
m^2_{\tilde q_1} \sin^2 \tilde\theta_q + m^2_{\tilde q_2} \cos^2
\tilde\theta_q - \hat m_q^2(Q_0) \, .
\label{eq:mlr}
\end{eqnarray}
The tree-level definition of the mixing angle $\tilde\theta_q$ in
Eq.~(\ref{eq:sqmix}) corresponds to the following counter term at NLO,
\begin{eqnarray}
\delta\tilde\theta_q & = & \frac{{\rm tg}~2\tilde\theta_q}{2} \left\{
\frac{\delta \hat m_q}{\hat m_q(Q_0)} + \frac{\delta \bar A_q}{\bar
A_q(Q_0)-\mu r_q} - \frac{\delta m_{\tilde q_1}^2 - \delta m_{\tilde
q_2}^2}{m_{\tilde q_1}^2 - m_{\tilde q_2}^2} \right\} \, ,
\label{eq:sqmixct1}
\end{eqnarray}
\begin{eqnarray}
\delta m_{\tilde q_{1/2}}^2 & = & -\Sigma_{11/22}(m_{\tilde q_{1/2}}^2)
\, .
\label{eq:sqmassct}
\end{eqnarray}
However, in order to avoid artificial singularities in physical
observables for stop/sbottom masses $m_{\tilde q_{1,2}}$ close to each
other, in our calculation the mixing angle of the squark fields has been
renormalized via the anti-Hermitian counter term\footnote{Alternatively
the mixing angle can be renormalized by the counter term $\delta
\theta_q = \Re e\Sigma_{12}(Q^2) /(m_{\tilde q_2}^2
- m_{\tilde q_1}^2)$ with $Q^2=m_{\tilde q_1}^2$ or $Q^2=m_{\tilde
  q_2}^2$ \cite{beenakker} which leads to a result free of the
artificial singularities for stop/sbottom masses $m_{\tilde q_{1,2}}$
close to each other, too. Another option free of these singularities is
provided by taking the residue of Eq.~(\ref{eq:sqmixct}) for equal
masses for the mixing angle counter term \cite{boudjema}.}
\cite{h2sqcorr,hsqimp},
\begin{eqnarray}
\delta \theta_q & = & \frac{1}{2} \frac{\Re e\Sigma_{12}(m_{\tilde
q_1}^2) + \Re e\Sigma_{12}(m_{\tilde q_2}^2)}{m_{\tilde q_2}^2
- m_{\tilde q_1}^2} \, ,
\label{eq:sqmixct}
\end{eqnarray}
where $\Sigma_{12}$ denotes the off-diagonal part of the stop/sbottom
self-energy (see Fig.~\ref{fg:sqself}) describing transitions from the
first to the second mass eigenstate or {\it vice versa},
\begin{eqnarray}
\Sigma_{12}(m^2) & = & - C_F\frac{\alpha_s}{\pi} \left\{ M_\sgl \hat
m_q(Q_0) B_0[m^2;M_\sgl,\hat m_q(Q_0)] \right. \nonumber \\
& & \left. \hspace*{2cm} + \frac{\sin 2\tilde\theta_q}{4} \left[
A_0(m_{\tilde q_2}) - A_0(m_{\tilde q_1}) \right] \right\} \cos
2\tilde\theta_q \, .
\label{eq:sig12}
\end{eqnarray}
This implies a finite shift $\Delta \tilde\theta_q$ to the mixing angle
$\tilde\theta_q$ of Eq.~(\ref{eq:sqmix}),
\begin{equation}
\theta_q = \tilde\theta_q + \Delta \tilde\theta_q \qquad , \qquad
\Delta \tilde\theta_q = \delta \tilde\theta_q - \delta \theta_q
\label{eq:sqmix1}
\end{equation}
which modifies the relations of Eq.~(\ref{eq:mlr}) by replacing $\tilde
\theta_q = \theta_q-\Delta \tilde \theta_q$.  The scale and scheme
dependence of the input parameters determining the squark pole masses is
explicitly compensated by the corrections $\Delta m^2_{\tilde q_{1,2}}$
to the tree-level relations of Eq.~(\ref{eq:sqmass}), while the scale
dependence of the mixing angle is compensated by the finite shift
$\Delta\tilde\theta_q$ in Eq.~(\ref{eq:sqmix1}). It should be noted that
the corrected mixing angle fulfills the relation $\sin^2 2\theta_q +
\cos^2 2\theta_q = 1$ at NLO consistently which will be necessary for
the cancellation of the ultraviolet divergences of the NLO corrections
to Higgs decays into squarks. The scale of the strong coupling constants
$\alpha_s$ in Eqs.~(\ref{eq:sigii}, \ref{eq:dmsq}, \ref{eq:daq},
\ref{eq:dmq}, \ref{eq:sig12}) has been identified with the input scale
$Q_0$.

In order to obtain fully consistent input parameters
Eqs.~(\ref{eq:deltab}, \ref{eq:sqmix}, \ref{eq:sqmass}, \ref{eq:mlr})
would have to be solved iteratively until the squark pole masses do not
change anymore.  Performing this iteration explicitly, however, shows
that for certain MSSM scenarios convergence cannot be reached. This
happens in particular for scenarios where the parameters $M^2_{\tilde
q_L}$ and $M^2_{\tilde q_R}$ are close to each other so that the mixing
angle $\tilde\theta_q$ is driven towards $\pm\pi/4$ and $\Delta
m_{\tilde q_{2}}^2$ towards $-\Delta m_{\tilde q_{1}}^2$. This leads to
a situation in which the squark pole masses do not change anymore, but
the mixing angles $\tilde\theta_q, \theta_q$ do not correspond to the
relations of Eqs.~(\ref{eq:sqmix}, \ref{eq:sqmix1}) so that the iterated
parameters are inconsistent. This situation can be avoided by stopping
the iteration earlier before running into this inconsistent limit. Since
the iteration adds contributions beyond NLO to the input parameters and
the squark masses, the iteration effects can be considered as arbitrary
in a NLO analysis.  We have chosen a procedure where no iteration is
performed as described above.  With the shifted parameters $\tilde
M^2_{\tilde q_{L,R}}$ of Eq.~(\ref{eq:mlr}) the squark pole masses are
calculated by Eq.~(\ref{eq:sqmass}) with vanishing $\Delta m_{\tilde
q_{1,2}}^2$ terms and the mixing angles according to
Eqs.~(\ref{eq:sqmix}, \ref{eq:sqmix1}), since the NLO corrections are
absorbed in the shifted values for $\tilde M_{\tilde q_{L,R}}$. However,
for a consistent treatment of the $\Delta_q$ terms of
Eq.~(\ref{eq:deltab}) iterations for the effective bottom mass $\hat
m_b(Q_0)$ are performed for the sbottom sector. This ensures that the
effective bottom mass of the corrected mass matrix corresponds to the
NLO sbottom pole masses consistently.

The NLO neutral Higgs couplings to squarks read in the current
eigenstate basis as
\begin{eqnarray}
g_{\tilde q_L \tilde q_L}^\Phi(\mu_R) & = & \hat m_q^2(\mu_R) g_1^\Phi +
M_Z^2 (I_{3q} - e_q\sin^2\theta_W) g_2^\Phi \nonumber \\
g_{\tilde q_R \tilde q_R}^\Phi(\mu_R) & = & \hat m_q^2(\mu_R) g_1^\Phi +
M_Z^2 e_q\sin^2\theta_W g_2^\Phi \nonumber \\
g_{\tilde q_L \tilde q_R}^\Phi(\mu_R) & = & -\frac{\hat m_q(\mu_R)}{2}
\left[ \mu g_3^\Phi - \bar A_q(\mu_R) g_4^\Phi \right]
\label{eq:hsbsbcoup}
\end{eqnarray}
with the couplings $g_i^\Phi$ listed in Table \ref{tb:hsfcoup}. Note
that we use a common renormalization scale $\mu_R$ for the effective
quark mass $\hat m_q$ and the trilinear coupling $\bar A_q$ for
simplicity. The corresponding couplings to the stop/sbottom mass
eigenstates $\tilde q_{1,2}$ are obtained by the appropriate rotations
according to Eq.~(\ref{eq:couprot}) by the on-shell mixing angle
$\theta_q$ of Eq.~(\ref{eq:sqmix1}).  The choice of the on-shell mixing
angle $\theta_q$ for these rotations implies that this is also the
relevant mixing angle involved in the physical processes. The on-shell
mixing angle is uniquely related to the soft SUSY-breaking input
parameters according to Eqs.~(\ref{eq:sqmix},\ref{eq:sqmix1}) and thus in
particular to the parameter $A_q$.

In our numerical analysis we include the running of the effective quark
mass $\hat m_q$ and the trilinear coupling $\bar A_q$ up to the NLL
level in (SUSY--)QCD. Note that due to the counter terms for the
effective quark masses of Eq.~(\ref{eq:dmq}) the heavy particles are
decoupled from its running.  We have included 5 (6) light flavours in
the NLL running of the effective bottom (top) masses. For the running of
the trilinear couplings $\bar A_q$ we have kept the contributions of all
coloured particles including the top quark, the squarks and the gluino
which is consistent with the counter term $\delta \bar A_q$ of
Eq.~(\ref{eq:daq}). The running effective quark mass for $N_F=5,6$ light
flavours is thus given by
\begin{eqnarray}
\hat m_q (\mu_R) & = & \hat m_q (m_q)
\,\frac{c[\alpha_{s}(\mu_R)/\pi]}{c[\alpha_{s}(m_q)/\pi]}
\nonumber \\
c(x) & = & \left(\frac{23}{6}\,x\right)^{\frac{12}{23}} \,
\left[1+\frac{3731}{3174}\,x\right] \qquad \mbox{for $q=b \quad (N_F=5)$}
\nonumber \\
c(x) & = & \left(\frac{7}{2}\,x\right)^{\frac{4}{7}} \,
\left[1+\frac{137}{98}\,x\right] \qquad \quad \; \mbox{for $q=t \quad
(N_F=6)$} \, ,
\label{eq:msbarev}
\end{eqnarray}
where $m_q$ denotes the corresponding quark pole mass. The NLL evolution
of the trilinear couplings $\bar A_q$ with $N_F=6$ quark and squark
flavours is determined as the solution of the corresponding two-loop
renormalization group equations (RGEs) with respect to the strong
coupling constant \cite{susyrest,rge} and can be expressed
as\footnote{Note that the RGEs have been transformed into the
$\overline{MS}$ scheme appropriately \cite{susyrest}. In the
$\overline{DR}$ scheme the coefficient $1/6$ is replaced by $7/6$, the
coefficient $16/27$ by $4/27$ and the $\overline{MS}$ gluino mass
$M_3(Q_0)$ by the corresponding $\overline{DR}$ gluino mass in
Eq.~(\ref{eq:aqnll}).}
\begin{eqnarray}
\bar A_q(\mu_R) & = & \bar A_q(Q_0) + M_3(Q_0) \left\{-\frac{16}{9}\,
\left[\frac{\alpha_{s,SUSY}(\mu_R)}{\alpha_{s,SUSY}(Q_0)}-1\right]
\left[1+\frac{1}{6}\,\frac{\alpha_{s,SUSY}(Q_0)}{\pi} \right] \right.
\nonumber \\
& & \left. \hspace*{3.5cm} -
\frac{16}{27}\,\frac{\alpha_{s,SUSY}(Q_0)}{\pi}
\left[\frac{\alpha_{s,SUSY}^2(\mu_R)}{\alpha_{s,SUSY}^2(Q_0)}-1\right]
\right\}
\label{eq:aqnll}
\end{eqnarray}
with the running strong coupling $\alpha_{s,SUSY}$ including all
quarks, squarks and gluinos,
\begin{equation}
\alpha_{s,SUSY}(\mu) = \frac{12\pi}{\displaystyle
9\log (\mu^2/\Lambda_{SUSY}^2)} \left\{ 1-\frac{14}{9}\,
\frac{\displaystyle \log\log (\mu^2/\Lambda_{SUSY}^2)}
{\displaystyle \log (\mu^2/\Lambda_{SUSY}^2)} \right\} \, ,
\label{eq:als_susy}
\end{equation}
where $\Lambda_{SUSY}$ is determined by the matching condition
\begin{equation}
\alpha_{s,SUSY}(Q_0) = \alpha_s(Q_0) \left\{ 1 +
\frac{\alpha_s(Q_0)}{\pi}\left[ \frac{1}{6}\log\frac{Q_0^2}{m_t^2} +
\frac{1}{2}\log\frac{Q_0^2}{M_{\sgl}^2} + \frac{1}{24} \sum_{\tilde q_i}
\log\frac{Q_0^2}{m_{\tilde q_i}^2} \right] \right\}
\end{equation}
between the strong coupling $\alpha_{s,SUSY}$ including all quarks,
squarks and gluinos and the low-energy QCD couplings $\alpha_s$ with 5
active light flavours. The $\overline{MS}$ mass $M_3(Q_0)$ of the
gluino can be obtained from the gluino pole mass $M_{\tilde g}$,
\begin{eqnarray}
M_3(Q_0) & = & M_{\sgl} \left\{ 1-\frac{\alpha_s(M_{\sgl})}{\pi} \left[ C_A +
\frac{3}{4}C_A \log\frac{Q_0^2}{M_{\sgl}^2} \right. \right. \nonumber \\
& & \hspace*{0.5cm} \left. \left. + \frac{1}{4} \sum_{q,i} \left(
B_1(M_{\sgl}^2;m_q,m_{\sq_i})+\frac{\Gamma(1+\epsilon)
(4\pi)^\epsilon}{2\epsilon} + \frac{1}{2}\log\frac{\bar\mu^2}{Q_0^2}
\right. \right. \right. \nonumber \\
& & \left. \left. \left. \hspace*{2.0cm} -(-1)^i \frac{m_q}{M_{\sgl}}
\sin 2\theta_q B_0(M_{\sgl}^2;m_q,m_{\sq_i}) \right) \right] \right\} \, ,
\end{eqnarray}
where $C_A=3$.  The sum is performed over all 3 (s)quark generations.
For the first two generations we neglected the quark masses in the
squark mass matrices so that the corresponding squark pole masses are
given by
\begin{eqnarray}
m^2_{\tilde q_{L/R}} & = & \tilde{\overline{M}}_{\tilde q_{L/R}}^2(Q_0)
+ C_F\frac{\alpha_s(Q_0)}{\pi} \left\{ M_{\sgl}^2
\log\frac{Q_0^2}{M_{\sgl}^2} + \frac{\tilde{\overline{M}}_{\tilde
q_{L/R}}^2(Q_0)}{2} \log\frac{M_{\sgl}^2}{\tilde{\overline{M}}_{\tilde
q_{L/R}}^2(Q_0)} \right. \nonumber \\
& & \left. \!\!\!\! + \frac{1}{2} \tilde{\overline{M}}_{\tilde
q_{L/R}}^2(Q_0) + \frac{3}{2} M_{\sgl}^2 + \frac{\left[
M_{\sgl}^2-\tilde{\overline{M}}_{\tilde q_{L/R}}^2(Q_0)
\right]^2}{2\tilde{\overline{M}}_{\tilde q_{L/R}}^2(Q_0)} \log\left|
\frac{M_{\sgl}^2-\tilde{\overline{M}}_{\tilde
q_{L/R}}^2(Q_0)}{M_{\sgl}^2} \right| \right\}
\end{eqnarray}

\section{Results at NLO} \label{sc:h2sqsq}
\begin{figure}[htb]
\begin{center}
\begin{picture}(130,90)(0,0)
\DashLine(0,50)(50,50){5}
\DashLine(50,50)(100,100){5}
\DashLine(50,50)(100,0){5}
\put(-15,48){$\Phi$}
\put(105,98){$\tilde q$}
\put(105,-2){$\overline{\tilde q}$}
\end{picture} \\
\caption{\label{fg:lodia} \it MSSM Higgs boson decays into
squark-antisquark pairs at leading order.}
\end{center}
\end{figure}
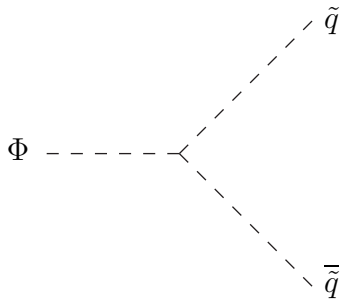
At leading order the MSSM Higgs boson decays into squark-antisquark
pairs are described by the diagram shown in Fig.~\ref{fg:lodia}. The
corresponding partial decay width reads as
\begin{equation}
\Gamma_{LO}(\Phi\to \tilde q_i \overline{\tilde q}_j) = \frac{N_c
G_F}{2\sqrt{2}\pi M_\Phi} \left[g_{\tilde q_i \tilde
q_j}^\Phi(\mu_R)\right]^2 \sqrt{\lambda_{ij}}
\end{equation}
with the two-body phase-space function
\begin{equation}
\lambda_{ij} = \left(1 - \frac{m_{\tilde q_i}^2}{M_\Phi^2} -
\frac{m_{\tilde q_j}^2}{M_\Phi^2} \right)^2 - 4\frac{m_{\tilde q_i}^2
m_{\tilde q_j}^2}{M_\Phi^4} \, .
\label{eq:lambda}
\end{equation}
While Higgs boson decays into squarks of the first two generations are
generally suppressed, Higgs decays into sbottoms and stops develop
sizable branching ratios, whenever they are kinematically possible
\cite{ohmann}. In particular scenarios their branching ratios can reach
a level of 80--90\% \cite{ohmann,h2sqcorr,spira:98}.

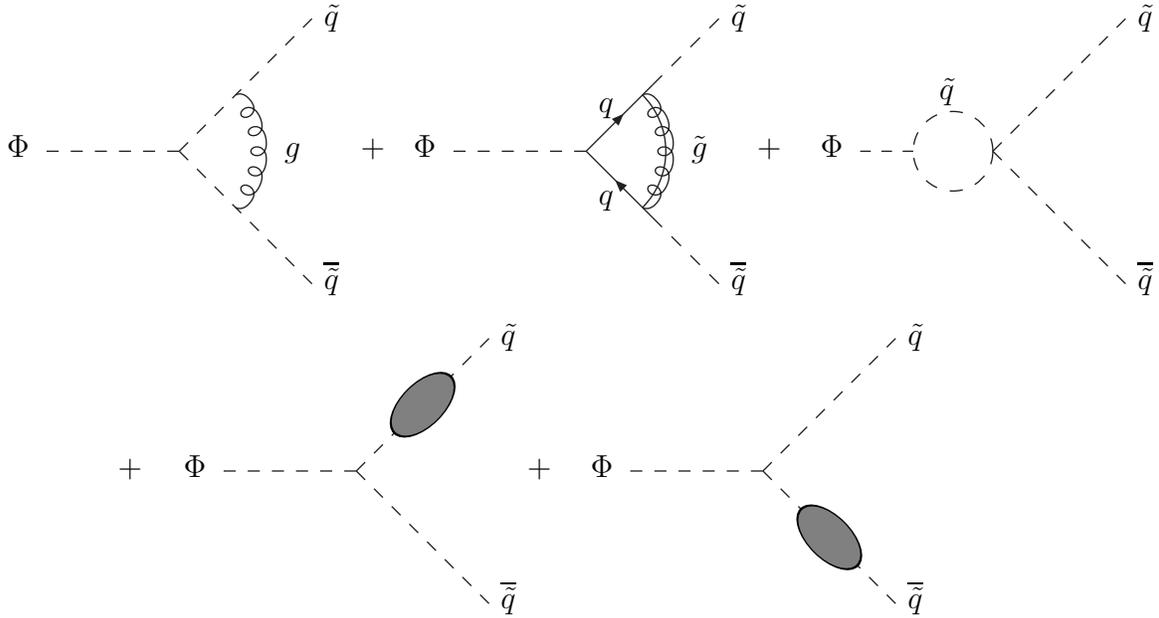
\begin{figure}[htb]
\begin{center}
\begin{picture}(130,90)(0,0)
\DashLine(0,50)(50,50){5}
\DashLine(50,50)(100,100){5}
\DashLine(50,50)(100,0){5}
\GlueArc(50,50)(30,-45,45){3}{5}
\put(-15,48){$\Phi$}
\put(105,98){$\tilde q$}
\put(105,-2){$\overline{\tilde q}$}
\put(90,48){$g$}
\end{picture}
\begin{picture}(130,120)(-20,0)
\DashLine(0,50)(50,50){5}
\DashLine(75,75)(100,100){5}
\DashLine(75,25)(100,0){5}
\ArrowLine(50,50)(75,75)
\CArc(50,50)(30,-45,45)
\GlueArc(50,50)(30,-45,45){3}{5}
\ArrowLine(75,25)(50,50)
\put(-15,48){$\Phi$}
\put(105,98){$\tilde q$}
\put(105,-2){$\overline{\tilde q}$}
\put(55,65){$q$}
\put(55,30){$q$}
\put(90,48){$\tilde g$}
\put(-35,48){$+$}
\put(115,48){$+$}
\end{picture}
\begin{picture}(130,120)(-40,0)
\DashLine(0,50)(20,50){5}
\DashLine(50,50)(100,100){5}
\DashLine(50,50)(100,0){5}
\DashCArc(35,50)(15,0,360){5}
\put(-15,48){$\Phi$}
\put(105,98){$\tilde q$}
\put(105,-2){$\overline{\tilde q}$}
\put(30,70){$\tilde q$}
\end{picture} \\
\begin{picture}(130,120)(0,0)
\DashLine(0,50)(50,50){5}
\DashLine(50,50)(100,100){5}
\DashLine(50,50)(100,0){5}
\GOval(75,75)(8,15)(45){0.5}
\put(-15,48){$\Phi$}
\put(105,98){$\tilde q$}
\put(105,-2){$\overline{\tilde q}$}
\put(-40,48){$+$}
\put(115,48){$+$}
\end{picture}
\begin{picture}(130,120)(-20,0)
\DashLine(0,50)(50,50){5}
\DashLine(50,50)(100,100){5}
\DashLine(50,50)(100,0){5}
\GOval(75,25)(8,15)(-45){0.5}
\put(-15,48){$\Phi$}
\put(105,98){$\tilde q$}
\put(105,-2){$\overline{\tilde q}$}
\end{picture} \\
\caption{\label{fg:nlodiav} \it Virtual corrections to MSSM Higgs boson
decays into squark-antisquark pairs at next-to-leading order. The
external self-energy blobs represent the self-energy diagrams of
Fig.~\ref{fg:sqself}.}
\end{center}
\end{figure}
\begin{figure}[htb]
\begin{center}
\begin{picture}(130,100)(0,0)
\DashLine(0,50)(50,50){5}
\DashLine(50,50)(100,100){5}
\DashLine(50,50)(100,0){5}
\Gluon(65,65)(100,65){3}{3}
\put(-15,48){$\Phi$}
\put(105,98){$\tilde q$}
\put(105,63){$g$}
\put(105,-2){$\overline{\tilde q}$}
\end{picture}
\begin{picture}(130,90)(-20,0)
\DashLine(0,50)(50,50){5}
\DashLine(50,50)(100,100){5}
\DashLine(50,50)(100,0){5}
\Gluon(65,35)(100,35){3}{3}
\put(-15,48){$\Phi$}
\put(105,98){$\tilde q$}
\put(105,33){$g$}
\put(105,-2){$\overline{\tilde q}$}
\end{picture} \\
\caption{\label{fg:nlodiar} \it Real corrections to MSSM Higgs boson
decays into squark-antisquark pairs at next-to-leading order.}
\end{center}
\end{figure}
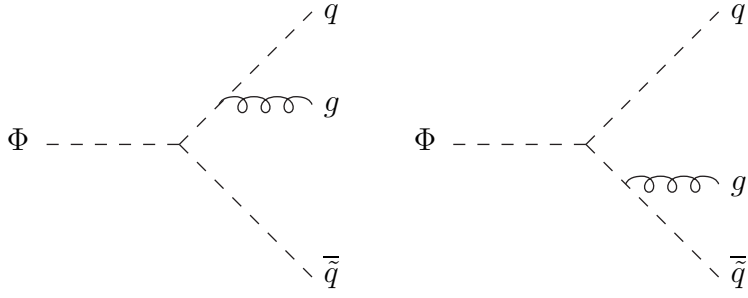
The SUSY--QCD corrections at NLO have been calculated in
Refs.~\cite{h2sqcorr}. However, the analytical results will be repeated
in this section in our notation and for the investigation of
calculational details once consistent input parameters at NLO are
included in the theoretical analysis. The SUSY--QCD corrections consist
of one-loop virtual corrections, mediated by the Feynman diagrams
depicted in Fig.~\ref{fg:nlodiav}, supplemented by the renormalization
of the mass, coupling and wave function parameters involved, and real
corrections due to gluon bremsstrahlung, see Fig.~\ref{fg:nlodiar}. We
have performed the loop and phase-space integration within dimensional
regularization in $n=4-2\epsilon$ dimensions so that ultraviolet and
infrared singularities appear as poles in $\epsilon$. The infrared
singularities cancel after adding the real and virtual corrections. The
ultraviolet poles disappear after adding the corresponding counter terms
which will be discussed in detail in the following, since they have to
correspond to the schemes in which the squark masses, mixing angles and
couplings are defined.

The wave-function counter terms $\delta Z_{ii}~(i=1,2)$ are determined
by normalizing the residues of the diagonalized squark propagators to
unity. The explicit calculation of the diagrams of Fig.~\ref{fg:sqself}
leads to the expressions
\begin{eqnarray}
\delta Z_{ii} & = & \frac{C_F}{2}\frac{\alpha_s}{\pi} \left\{ B_0(m_{\tilde
q_i}^2;0,m_{\tilde q_i}) - B_0(m_{\tilde q_i}^2;M_\sgl,m_q) +
2m_{\tilde q_i}^2B'_0(m_{\tilde q_i}^2;0,m_{\tilde q_i}) \right.
\nonumber \\
& & \left. +\left[M_\sgl^2+m_q^2-m_{\tilde q_i}^2 + (-1)^i 2M_\sgl m_q
\sin 2\theta_q \right] B'_0(m_{\tilde q_i}^2;M_\sgl,m_q) \right\} \, ,
\end{eqnarray}
where the derivative of the two-point function is defined as
\begin{equation}
B'_0(p^2;m_1,m_2) = \frac{\partial}{\partial p^2} B_0(p^2;m_1,m_2) \, .
\end{equation}
Throughout our calculation in this section the quark mass $m_q$ is
defined as $\hat m_b(\mu_R)$ of Eq.~(\ref{eq:deltab}) for decays into
sbottom pairs and as $\hat m_t(\mu_R)$ of Eq.~(\ref{eq:tmass}) for
decays into stops.  The mixing angle is renormalized by the
anti-Hermitian counter term of Eq.~(\ref{eq:sqmixct}). The trilinear
couplings $\bar A_q(\mu_R)$ ($q=b,t$) are defined in the $\overline{MS}$
scheme at the scale $\mu_R$ which is identical to the running couplings
in the $\overline{DR}$ definition at NLO. Thus, the counter term of
$\bar A_q(\mu_R)$ can be cast into the form
\begin{equation}
\delta \bar A_q = C_F \frac{\alpha_s}{\pi} \Gamma(1+\epsilon)
(4\pi)^\epsilon M_\sgl \left\{ \frac{1}{\epsilon} + \log \frac{\bar
\mu^2}{\mu_R^2} \right\} \, .
\label{eq:daqp}
\end{equation}
The quark mass counter term $\delta m_q$ is given by
\begin{equation}
\delta m_q = \delta \hat m_q
\end{equation}
with the top and bottom mass counter terms $\delta \hat m_q$ of
Eq.~(\ref{eq:dmq}) with $Q=\mu_R$.

The calculation of the diagrams of Fig.~\ref{fg:nlodiav} for the virtual
corrections and those of Fig.~\ref{fg:nlodiar} for the real corrections
leads to the final result after renormalization,
\begin{eqnarray}
\Gamma(\Phi\to \sq_i \sqb_j) & = & \Gamma_{LO}(\Phi\to \sq_i \sqb_j)
\left\{ 1+C^\Phi \frac{\alpha_s}{\pi}\right\} \nonumber \\
C^\Phi & = & \frac{C_F}{2} [C^\Phi_1 + C^\Phi_2 + C^\Phi_3] + C^\Phi_{CT} +
C^\Phi_{real} \nonumber \\
C^\Phi_1 & = & B_0(m_{\tilde q_i}^2;0,m_{\tilde q_i}) + B_0(m_{\tilde
q_j}^2;0,m_{\tilde q_j}) - B_0(M_\Phi^2; m_{\tilde q_i}, m_{\tilde q_j})
\nonumber \\
& & - 2(M_\Phi^2-m_{\tilde q_i}^2-m_{\tilde q_j}^2) C_0(m_{\tilde
q_i}^2,M_\Phi^2,m_{\tilde q_j}^2;0,m_{\tilde q_i},m_{\tilde q_j})
\nonumber \\
C^{h,H}_2 & = & \frac{g_q^{h,H}}{g_{\tilde q_i\tilde q_j}^{h,H}} \left\{
\delta_{ij} m_q^2 \left[ B_0(m_{\tilde q_i}^2;M_\sgl,m_q) +
B_0(m_{\tilde q_j}^2;M_\sgl,m_q) +2B_0(M_{h,H}^2;m_q,m_q) \right.
\right. \nonumber \\
& & \left. \left. + (2M_\sgl^2+2m_q^2-m_{\tilde q_i}^2-m_{\tilde q_j}^2)
C_0(m_{\tilde q_i}^2,M_{h,H}^2,m_{\tilde q_j}^2;M_\sgl, m_q, m_q)
\right] \right. \nonumber \\
& & +{\cal R}_{ij} M_\sgl m_q \left[ (M_{h,H}^2-4m_q^2) C_0(m_{\tilde
q_i}^2,M_{h,H}^2,m_{\tilde q_j}^2;M_\sgl, m_q,m_q) \right. \nonumber \\
& & \left. \left. -B_0(m_{\tilde q_i}^2;M_\sgl,m_q) - B_0(m_{\tilde
q_j}^2;M_\sgl,m_q) \right] \right\} \nonumber \\
C^A_2 & = & \frac{g_q^A}{g_{\tilde q_i\tilde q_j}^A} \left\{ m_q^2 \sin
2\theta_q \left[ B_0(m_{\tilde q_i}^2;M_\sgl,m_q) - B_0(m_{\tilde
q_j}^2;M_\sgl,m_q) \right.  \right. \nonumber \\
& & \left. \left. + (m_{\tilde q_i}^2-m_{\tilde q_j}^2) C_0(m_{\tilde
q_i}^2,M_A^2,m_{\tilde q_j}^2;M_\sgl, m_q, m_q) \right] \right.
\nonumber \\
& & \left. -(-1)^i M_\sgl m_q \left[ M_A^2 C_0(m_{\tilde
q_i}^2,M_A^2,m_{\tilde q_j}^2;M_\sgl, m_q,m_q) \right. \right. \nonumber
\\
& & \left. \left. -B_0(m_{\tilde q_i}^2;M_\sgl,m_q) - B_0(m_{\tilde
q_j}^2;M_\sgl,m_q) \right] \right\} \nonumber \\
C^{h,H}_3 & = & -\frac{{\cal S}_{ik} g^{h,H}_{\tilde q_k\tilde q_l}
{\cal S}_{lj}}{g_{\tilde q_i\tilde q_j}^{h,H}} B_0(M_{h,H}^2;m_{\tilde
q_k}, m_{\tilde q_l}) \nonumber \\
C^A_3 & = & B_0(M_A^2;m_{\tilde q_i}, m_{\tilde q_j}) \nonumber \\
\frac{\alpha_s}{\pi} C^{h,H}_{CT} & = & \delta Z_{ii}+\delta Z_{jj} +
\frac{2}{g_{\tilde q_i\tilde q_j}^{h,H}} \left[ \frac{\partial
g^{h,H}_{\tilde q_i\tilde q_j}}{\partial m_q} \delta m_q +
\frac{\partial g^{h,H}_{\tilde q_i\tilde q_j}}{\partial A_q} \delta \bar A_q
+ {\cal T}_{ij} \Delta\theta_q \right] \nonumber \\
\frac{\alpha_s}{\pi} C^A_{CT} & = & \delta Z_{ii}+\delta Z_{jj} +
\frac{2}{g_{\tilde q_i\tilde q_j}^A} \left[ \frac{\partial g^A_{\tilde
q_i\tilde q_j}}{\partial m_q} \delta m_q + \frac{\partial g^A_{\tilde
q_i\tilde q_j}}{\partial A_q} \delta \bar A_q \right] \nonumber \\
C^\Phi_{real} & = & C_F \Gamma(1+\epsilon) \left(
\frac{4\pi\bar\mu^2}{M_\Phi^2}\right)^\epsilon \left\{
\frac{1-\rho_i-\rho_j}{2\beta} \left[ \frac{\log x_0}{\epsilon} +
\log x_0 \log\frac{\rho_i\rho_j x_0}{\beta^4}
\right. \right. \nonumber \\
& & \left. \left. - \frac{1}{2} \log^2 x_1 - \frac{1}{2} \log^2 x_2 + 4
Li_2\left(\frac{1-x_0}{-x_0}\right) - 2 Li_2(1-x_1) - 2 Li_2(1-x_2) \right]
\right. \nonumber \\
& & \left. + \frac{1}{\epsilon} +
\log\frac{\rho_i\rho_j}{\beta^4} + 4 - \frac{1+\rho_i+\rho_j}{2\beta}
\log x_0 + \frac{\rho_i\log x_1 + \rho_j\log x_2}{\beta}\right\} \, ,
\label{eq:cphi}
\end{eqnarray}
where in $C_{real}$ we used the abbreviations $\beta =
\sqrt{\lambda_{ij}}$ with the two-body phase-space function
$\lambda_{ij}$ of Eq.~(\ref{eq:lambda}) and
\begin{equation}
x_0 = \frac{1-\rho_i-\rho_j-\beta}{1-\rho_i-\rho_j+\beta} \quad , \quad
x_1 = \frac{1-\rho_i+\rho_j-\beta}{1-\rho_i+\rho_j+\beta} \quad , \quad
x_2 = \frac{1+\rho_i-\rho_j-\beta}{1+\rho_i-\rho_j+\beta}
\end{equation}
with $\rho_i = m_{\tilde q_i}^2/M_\Phi^2$.  The two mixing matrices
${\cal R}$ and ${\cal S}$ used in the coefficients $C^{h,H}_2$ and
$C^{h,H}_3$ are given by
\begin{equation}
{\cal R} = \left( \begin{array}{lr} \sin 2\theta_q & \cos 2\theta_q \\
\cos 2\theta_q & -\sin 2\theta_q \end{array} \right) \quad , \quad
{\cal S} = \left( \begin{array}{lr} \cos 2\theta_q & -\sin 2\theta_q \\
-\sin 2\theta_q & -\cos 2\theta_q \end{array} \right) \, .
\end{equation}
The three-point function $C_0$ is defined as \cite{passvelt}
\begin{equation}
C_0(p_1^2,p_2^2,p_{12}^2;m_1,m_2,m_3) = \int \frac{d^n k}{(2\pi)^n}
\frac{-i(4\pi)^2
\bar\mu^{2\epsilon}}{[k^2-m_1^2][(k+p_1)^2-m_2^2][(k+p_{12})^2-m_3^2]}
\end{equation}
with $p_{12} = p_1+p_2$. The finite remainders of the mixing angle
renormalization at the external squark legs can be cast into the form of
the matrix
\begin{eqnarray}
{\cal T} & = & \left( \begin{array}{cc} 2g_{\tilde q_1\tilde q_2}^\Phi &
g_{\tilde q_1\tilde q_1}^\Phi + g_{\tilde q_2\tilde q_2}^\Phi \\
g_{\tilde q_1\tilde q_1}^\Phi + g_{\tilde q_2\tilde q_2}^\Phi &
2g_{\tilde q_1\tilde q_2}^\Phi \end{array} \right)
\end{eqnarray}
multiplied by the finite shift
\begin{equation}
\Delta\theta_q = \frac{\Re e \Sigma_{12}(m_{\tilde q_2}^2)}{m_{\tilde
q_2}^2-m_{\tilde q_1}^2} - \delta\theta_q = \frac{1}{2} \Re e
\frac{\Sigma_{12}(m_{\tilde q_2}^2)-\Sigma_{12}(m_{\tilde
q_1}^2)}{m_{\tilde q_2}^2-m_{\tilde q_1}^2} \, ,
\label{eq:deltath}
\end{equation}
which includes the corresponding anti-Hermitian counter term
$\delta\theta_q$ of the mixing angle as given in Eq.~(\ref{eq:sqmixct}).
In Eq.~(\ref{eq:deltath}) the singularity for $m_{\tilde q_2}^2\to
m_{\tilde q_1}^2$ is cancelled by the anti-Hermitian counter term.
Finally the derivatives of the squark couplings $g_{\tilde q_i\tilde
q_j}^\Phi$ to the Higgs bosons are given by
\begin{eqnarray}
\frac{\partial g_{\tilde q_i\tilde q_j}^{h,H}}{\partial m_q} & = & 2m_q
g_q^{h,H} \delta_{ij} + \frac{g_{\tilde q_L\tilde q_R}^{h,H}}{m_q} {\cal
R}_{ij} \nonumber \\
\frac{\partial g_{\tilde q_i\tilde q_j}^{h,H}}{\partial A_q} & = &
\frac{m_q}{2} g_q^{h,H}{\cal R}_{ij} \nonumber \\
\frac{\partial g_{\tilde q_1\tilde q_2}^A}{\partial m_q} & = &
-\frac{\partial g_{\tilde q_2\tilde q_1}^A}{\partial m_q} =
\frac{g_{\tilde q_1\tilde q_2}^A}{m_q} \nonumber \\
\frac{\partial g_{\tilde q_1\tilde q_2}^A}{\partial A_q} & = &
-\frac{\partial g_{\tilde q_2\tilde q_1}^A}{\partial A_q} =
\frac{m_q}{2} g_q^A \, .
\end{eqnarray}
The results for the coefficients $C^\Phi$ are ultraviolet and infrared
finite after all individual contributions are added up since our
renormalized mixing angles fulfill the relation $\sin^22\theta_q +
\cos^22\theta_q=1$ consistently. The scale of $\alpha_s$ in
Eq.(\ref{eq:cphi}) is identified with the renormalization scale
everywhere apart from the $\Delta_b$ term within the effective bottom
mass of Eq.~(\ref{eq:deltab}). It should be
noted that our final results do not contain any leading $\tgb$-enhanced
corrections of the $\Delta_b$-type anymore due to our scheme choices of
the effective masses and couplings at NLO. Due to this property the
problems of a naive renormalization program have been solved
consistently leading to moderate radiative corrections to the Higgs
decay widths into squark-antisquark pairs as will be shown explicitly
in the next section.

By using power counting arguments along the lines of Ref.~\cite{guasch}
it can be shown that our approach resums all leading terms of ${\cal
O}(\alpha_s^n\mu^n{\rm tg}^n \beta)$ so that the residual HO corrections
are free of these leading terms. In the current eigenstate basis there
are three sources of terms proportional to $\tgb$: {\it (i)} the Higgs
couplings to squarks and quarks as in Table \ref{tb:hcoup} and
Eq.~(\ref{eq:hsbsbcoup}) which drop out in the relative corrections;
{\it (ii)} off-diagonal mass insertions in the sbottom propagators which
are power-suppressed by $m_b/M_S$ due to the KLN theorem and the
$m_b/M_S$ structure of these insertions; (iii) potential
$\tgb$-enhancements of the counter terms. The $\overline{MS}$
renormalization of $A_b$ as in Eq.~(\ref{eq:daqp}) does not involve any
$\tgb$ enhanced corrections as can also be inferred from the beta
functions of the RGEs for the trilinear couplings up to three-loop order
\cite{rge}. The $\tgb$ enhancement of the bottom-mass counter term
entering the corresponding bottom Yukawa couplings involved in the Higgs
couplings to bottom quarks and the sbottom mass matrix is explicitly
absorbed in the effective bottom mass of Eq.~(\ref{eq:deltab}) which
contains the corresponding threshold correction $\Delta_b$ in resummed
form \cite{deltamb,guasch}. Therefore the counter terms $\delta m_b$ and
$\delta \bar A_b$ are free of $\tgb$ enhanced corrections at all orders
up to terms which are suppressed by $m_b/M_S$.

Close to the threshold $M_\Phi\sim m_{\tilde q_i} + m_{\tilde q_j}$ the
NLO SUSY--QCD corrections develop Coulomb singularities,
\begin{equation}
C^\Phi \to \frac{C_F}{2}
\frac{\pi^2}{\sqrt{\lambda_{ij}}} \frac{4 m_{\tilde
q_i} m_{\tilde q_j}}{(m_{\tilde q_i} + m_{\tilde q_j})^2} \qquad
\mbox{for}~M_\Phi \to m_{\tilde q_i} + m_{\tilde q_j} \, ,
\end{equation}
which agrees with the usual Sommerfeld rescattering correction factor
\cite{sommerfeld}. These singularities for $\sqrt{\lambda_{ij}}\to 0$
can be regularized by taking into account the finite decay widths of the
squarks. Moreover, these Coulomb factors can be resummed systematically.
Both effects are expected to be relevant close to threshold, but are
beyond the scope of this paper.

On the other hand far above the threshold $M_\Phi \gg m_{\tilde q_{i,j}}
\gg m_q$ the NLO corrections approach the asymptotic limit
\begin{eqnarray}
C^\Phi & \to & C_F \left\{ 2\frac{m_q}{g_{\tilde q_i\tilde q_j}^\Phi}
\frac{\partial g_{\tilde q_i\tilde q_j}^\Phi}{\partial m_q} \left[
\frac{3}{4} \log\frac{\mu_R^2}{M_\Phi^2} - \frac{1}{4}\log
\frac{M_\Phi^2}{M_S^2} + \frac{7}{4} + \gamma_q
\right] \right. \nonumber \\
& & \left. \hspace*{0.5cm} + 2\frac{M_S}{g_{\tilde q_i\tilde
q_j}^\Phi} \frac{\partial g_{\tilde q_i\tilde q_j}^\Phi}{\partial A_q}
\left[ -\log\frac{\mu_R^2}{M_S^2} + \frac{1}{4}
\log^2\frac{M_\Phi^2}{M_S^2} + \frac{\zeta_2}{2} - 2 \right] + \delta_\Phi
\right\} \nonumber \\
\delta_{h,H} & = & \left[ \frac{m_q}{g_{\tilde q_i\tilde q_j}^\Phi}
\frac{\partial g_{\tilde q_i\tilde q_j}^\Phi}{\partial m_q} - 1 \right]
\left[ \log\frac{M_\Phi^2}{M_S^2} - 2 \right] +
\frac{g^\Phi_q m_q^2}{g_{\tilde q_i\tilde q_j}^{h,H}} \delta_{ij}
\log\frac{M_\Phi^2}{M_S^2} \nonumber \\
& & +~{\cal T}_{ij}~\frac{m_q}{M_S} \frac{\cos
2\theta_q}{g_{\tilde q_i\tilde q_j}^\Phi} \left[
1-\frac{1}{2}\log\frac{M_S^2}{m_q^2} \right] \nonumber \\
\delta_A & = & 0 \, ,
\label{eq:largemh}
\end{eqnarray}
where we have identified all supersymmetric masses, $M_S = m_{\tilde
q_1} = m_{\tilde q_2} = M_{\tilde g}$, and neglected the quark mass
against the other masses. The term $\gamma_q$ is given by
\begin{equation}
\gamma_t = \frac{A_t-\mu / \tgb}{4M_S} \, ,\qquad
\gamma_b = \frac{A_b}{4M_S}
\end{equation}
thus confirming the absence of large corrections of ${\cal
O}(\alpha_s\mu\tgb)$\footnote{In the on-shell scheme, which relates the
renormalization of $A_q$ to the on-shell quark and squark mass as well
as the mixing angle counter terms as in Eq.~(\ref{eq:sqmixct1}) and
renormalizes the quark mass on-shell, large corrections of ${\cal
O}(\alpha_s\mu^2{\rm tg}^2\beta)$ emerge  for the partial Higgs decay
widths into squarks.}. The asymptotic result far above the threshold
develops a double logarithmic contribution in the second square bracket
of Eq.~(\ref{eq:largemh}) which originates from the second diagram of
Fig.~\ref{fg:nlodiav} in the Sudakov limit. It cannot be absorbed in an
appropriate choice of the renormalization scale $\mu_R$ but will cancel
against the corresponding double logarithm of gluino radiation in
association with quark-squark pairs in the large Higgs mass limit which
has not been taken into account in this work. In the following we will
identify $\mu_R$ with the corresponding Higgs mass $M_\Phi$ as the
central scale choice.

\section{Numerical Results} \label{sc:numerics}
The numerical analysis of the neutral Higgs boson decays into stop and
sbottom pairs is performed for two MSSM scenarios, one close to the one
of Ref.~\cite{hsqimp} where we lifted the gluino and squark masses of
the first two generations beyond the mass bounds from the LHC
\cite{squarklhc}, and a second scenario with large SUSY-breaking masses
and large mixing in the stop sector as representative cases:
\begin{eqnarray}
\mbox{{\bf A)}} \quad && Q_0 = 300~{\rm GeV}, \nonumber \\
&& \tgb = 30,\quad \overline{M}_{\tilde t_L}(Q_0) = \overline{M}_{\tilde
b_L}(Q_0) = \overline{M}_{\tilde \tau_{L,R}}(Q_0) = 300~{\rm GeV},
\nonumber \\
& & \overline{M}_{\tilde t_R}(Q_0) = 270~{\rm GeV}, \quad
\overline{M}_{\tilde b_R}(Q_0) = 330~{\rm GeV}, \nonumber \\
& & \overline{M}_{\tilde q_{L,R}}(Q_0) = \overline{M}_{\tilde
\ell_{L,R}}(Q_0) = M_{\tilde g} = 1~{\rm TeV}, \quad
\bar A_t(Q_0) = 150~{\rm GeV}, \nonumber \\
& & \bar A_b(Q_0) = -700~{\rm GeV},
\quad \bar A_\tau (Q_0) = 1~{\rm TeV}, \quad
M_2 = 1~{\rm TeV}, \quad \mu = 260~{\rm GeV} \nonumber \\ \nonumber
\\
\mbox{{\bf B)}} \quad && Q_0 = 500~{\rm GeV}, \nonumber \\
&&\tgb = 30,\quad
\overline{M}_{\tilde t_{L,R}}(Q_0) = \overline{M}_{\tilde b_{L,R}}(Q_0)
= \overline{M}_{\tilde \tau_{L,R}}(Q_0) = 500~{\rm GeV}, \nonumber \\
&& \overline{M}_{\tilde q_{L,R}}(Q_0) = \overline{M}_{\tilde
\ell_{L,R}}(Q_0) = M_{\tilde g} = 1~{\rm TeV}, \quad \bar A_t(Q_0) =
\bar A_b(Q_0) = -1.5~{\rm TeV}, \nonumber \\
&& \bar A_\tau (Q_0) = 0, \quad M_2 = 500~{\rm GeV}, \quad \mu =
500~{\rm GeV} \, ,
\end{eqnarray}
where $\overline{M}_{\tilde q_{L,R}}(Q_0)$ and $\overline{M}_{\tilde
\ell_{L,R}}(Q_0)$ denote the squark and slepton mass
parameters of the first two generations.  The results of this work have
been implemented in the program HDECAY \cite{hdecay} which calculates
the MSSM Higgs decay widths and branching ratios including the relevant
higher-order corrections \cite{spira:98}.  We use the RG-improved
two-loop expressions for the Higgs masses and couplings of
Ref.~\cite{rgi} which yield predictions for the Higgs boson masses that
agree with the diagrammatic calculations of Ref.~\cite{mssmrad} within
3--4\% in general. Thus the leading one- and two-loop corrections have
been included in the Higgs masses and the effective mixing angle
$\alpha$.  Consistency of our scheme with the scheme and scale choices
of Ref.~\cite{rgi} requires the evolution of our sbottom parameters
$\bar A_b$ and $\overline{M}_{\tilde b_{L/R}}$ to the scale
\begin{equation}
Q_b = \max\left\{ \sqrt{\overline{M}^2_{\tilde b_L}(Q_0) +
\overline{m}_b^2(m_t)}, \sqrt{\overline{M}^2_{\tilde b_R}(Q_0) +
\overline{m}_b^2(m_t)} \right\}
\label{eq:q_b}
\end{equation}
and the stop parameters $\bar A_t$ and $\overline{M}_{\tilde
t_{L/R}}$ to the scale
\begin{equation}
Q_t = \max\left\{ \sqrt{\overline{M}^2_{\tilde t_L}(Q_0) +
\overline{m}_t^2(m_t)},
\sqrt{\overline{M}^2_{\tilde t_R}(Q_0) + \overline{m}_t^2(m_t)} \right\}
\, .
\label{eq:q_t}
\end{equation}
Since the scales $Q_0$ and $Q_{b/t}$ are of similar order of magnitude
we neglect resummation effects so that the relations are given
by\footnote{Note that the left-handed soft supersymmetry-breaking squark
mass parameters $\overline{M}_{\tilde b_L}(Q_b)$ and
$\overline{M}_{\tilde t_L}(Q_t)$ are not equal, because they are
evaluated for different scales $Q_t$ and $Q_b$. We have modified the
calculation of Ref.~\cite{rgi} to account for these differences
consistently.}
\begin{eqnarray}
\bar A_b(Q_b) & = & \bar A_b(Q_0) + \left\{ C_F
\frac{\alpha_s(Q_0)}{\pi} M_3(Q_0) + \frac{3}{2} \frac{\alpha_t}{\pi}
\bar A_t(Q_0) + \frac{\alpha_b}{4\pi} \bar A_b(Q_0) \right\}
\log\frac{Q_b^2}{Q_0^2} \nonumber \\
\overline{M}^2_{\tilde b_L}(Q_b) & = & \overline{M}^2_{\tilde b_L}(Q_0)
+ \left\{ - C_F \frac{\alpha_s(Q_0)}{\pi} M_3^2(Q_0) + \frac{1}{4} \left[
\frac{\alpha_t}{\pi} X_t + \frac{\alpha_b}{\pi} X_b \right] \right\}
\log\frac{Q_b^2}{Q_0^2} \nonumber \\
\overline{M}^2_{\tilde b_R}(Q_b) & = & \overline{M}^2_{\tilde b_L}(Q_0)
+ \left\{ - C_F \frac{\alpha_s(Q_0)}{\pi} M_3^2(Q_0) +
\frac{\alpha_b}{4\pi} X_b \right\} \log\frac{Q_b^2}{Q_0^2} \nonumber \\
\bar A_t(Q_t) & = & \bar A_t(Q_0) + \left\{ C_F
\frac{\alpha_s(Q_0)}{\pi} M_3(Q_0) + \frac{\alpha_t}{4\pi}
\bar A_t(Q_0) + \frac{3}{2} \frac{\alpha_b}{\pi}
\bar A_b(Q_0) \right\} \log\frac{Q_t^2}{Q_0^2} \nonumber \\
\overline{M}^2_{\tilde t_L}(Q_t) & = & \overline{M}^2_{\tilde t_L}(Q_0)
+ \left\{ - C_F \frac{\alpha_s(Q_0)}{\pi} M_3^2(Q_0) + \frac{1}{4} \left[
\frac{\alpha_t}{\pi} X_t + \frac{\alpha_b}{\pi} X_b \right] \right\}
\log\frac{Q_t^2}{Q_0^2} \nonumber \\
\overline{M}^2_{\tilde t_R}(Q_t) & = & \overline{M}^2_{\tilde t_L}(Q_0)
+ \left\{ - C_F \frac{\alpha_s(Q_0)}{\pi} M_3^2(Q_0) +
\frac{\alpha_t}{4\pi} X_t \right\} \log\frac{Q_t^2}{Q_0^2}
\label{eq:transform}
\end{eqnarray}
with the abbreviations
\begin{eqnarray}
\alpha_b & = & \frac{\overline{m}_b^2(Q_0)}{2\pi v^2 \cos^2\beta} \nonumber \\
\alpha_t & = & \frac{\overline{m}_t^2(Q_0)}{2\pi v^2 \sin^2\beta} \nonumber \\
X_b & = & \overline{M}^2_{\tilde b_L}(Q_0) + \overline{M}^2_{\tilde
b_R}(Q_0) + M^2_{H_1} + \bar A_b^2(Q_0) \nonumber \\
X_t & = & \overline{M}^2_{\tilde t_L}(Q_0) + \overline{M}^2_{\tilde
t_R}(Q_0) + M^2_{H_2} + \bar A_t^2(Q_0)
\end{eqnarray}
and the soft SUSY-breaking Higgs mass parameters
\begin{eqnarray}
M^2_{H_1} & = & M_A^2 \sin^2\beta - \frac{M_Z^2}{2} \cos 2\beta - \mu^2
\nonumber \\
M^2_{H_2} & = & M_A^2 \cos^2\beta + \frac{M_Z^2}{2} \cos 2\beta - \mu^2
\, .
\end{eqnarray}
The top pole mass has been taken as $m_t=172.6$ GeV, while the bottom
quark pole mass has been chosen to be $m_b=4.60$~GeV, which corresponds
to a $\overline{MS}$ mass $\overline{m}_b(\overline{m}_b)=4.26$~GeV. The
strong coupling constant has been normalized to $\alpha_s(M_Z)=0.118$
which corresponds to the QCD scale $\Lambda_5=226.2$ MeV for 5 active
flavours. The related SUSY--QCD scale $\Lambda_{SUSY}$ of
Eq.~(\ref{eq:als_susy}) amounts to $\Lambda_{SUSY}= 2.028$ keV in
scenario A and $\Lambda_{SUSY}= 1.756$ keV in scenario B. The
$\overline{MS}$ masses involved in the scales $Q_{t,b}$ can be derived
as $\overline{m}_t(m_t) = 165.1$~GeV and $\overline{m}_b(m_t) =
2.78$~GeV while the $\overline{MS}$ gluino mass at the input scale $Q_0$
amounts to $M_3(Q_0)= 1.028$ TeV in scenario A and $M_3(Q_0)= 1.006$ TeV
in scenario B. The squark masses in the two scenarios amount in
particular to
\begin{eqnarray}
\mbox{\bf A)}~m_{\tilde t_1} = 165.5~{\rm
GeV},\quad m_{\tilde t_2} = 278.0~{\rm GeV},\quad m_{\tilde b_1} =
182.5~{\rm GeV},\quad m_{\tilde b_2} = 265.6~{\rm GeV}, \nonumber \\
\phantom{\mbox{\bf A)}} m_{\tilde u_L} = 989.8~{\rm
GeV},\quad\! m_{\tilde u_R} = 990.7~{\rm GeV},\quad\! m_{\tilde d_L} =
993.1~{\rm GeV},\quad\! m_{\tilde d_R} = 991.6~{\rm GeV} \nonumber \\
\mbox{\bf B)}~m_{\tilde t_1} = 202.1~{\rm GeV},\quad
m_{\tilde t_2} = 708.0~{\rm GeV},\quad m_{\tilde b_1} = 473.0~{\rm
GeV},\quad m_{\tilde b_2} = 536.3~{\rm GeV}, \nonumber \\
m_{\tilde u_L} = 1.011~{\rm
TeV},\quad\! m_{\tilde u_R} = 1.012~{\rm TeV},\quad\! m_{\tilde d_L} =
1.014~{\rm TeV},\quad\! m_{\tilde d_R} = 1.012~{\rm TeV} \, ,
\end{eqnarray}
where $m_{\tilde u_{L/R}}$ and $m_{\tilde d_{L/R}}$ denote the up- and
down-type squark masses of the first two generations.

\begin{figure}[hbtp]
\begin{picture}(100,500)(0,0)
\put(40.0,120.0){\includegraphics{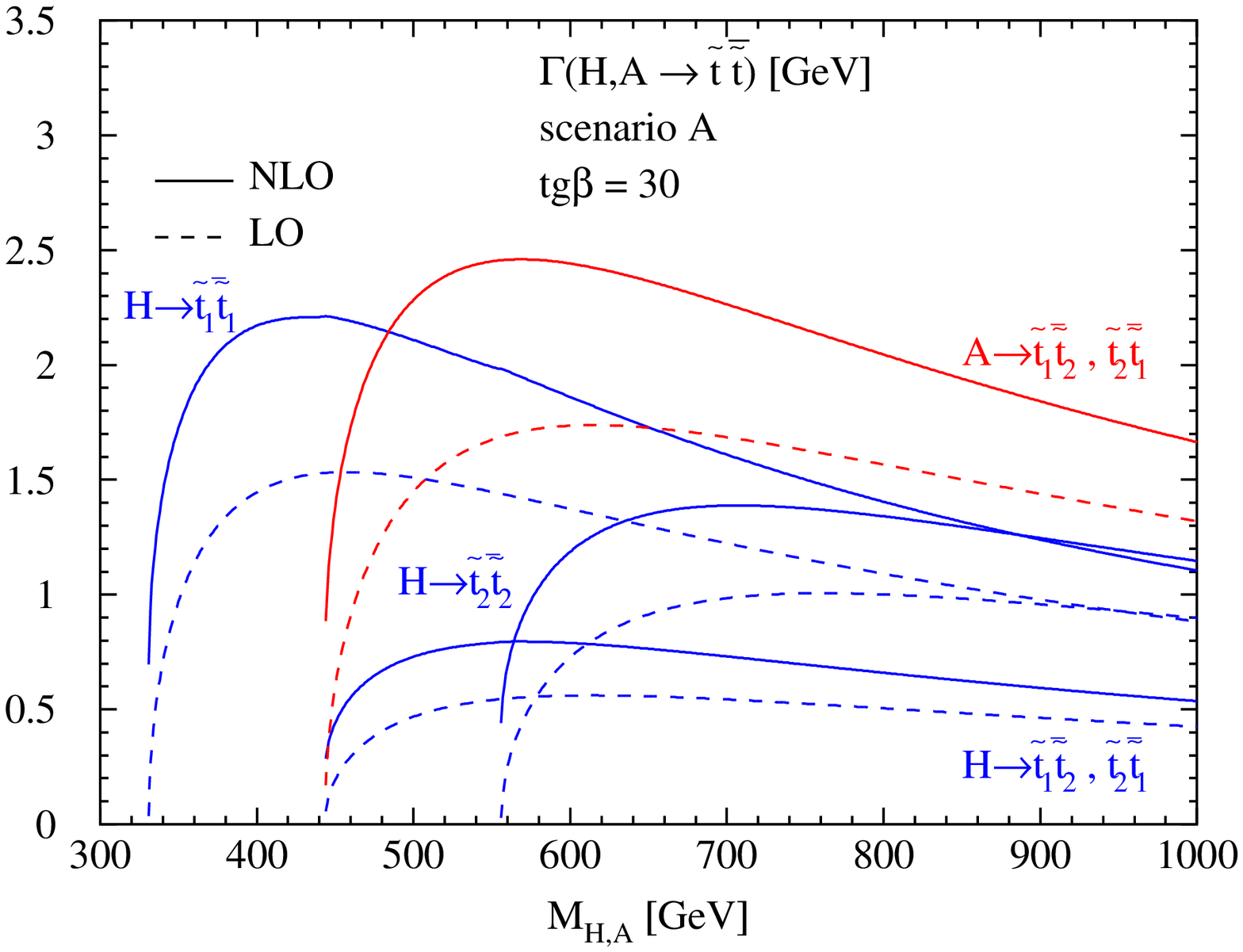}}
\put(40.0,-135.0){\includegraphics{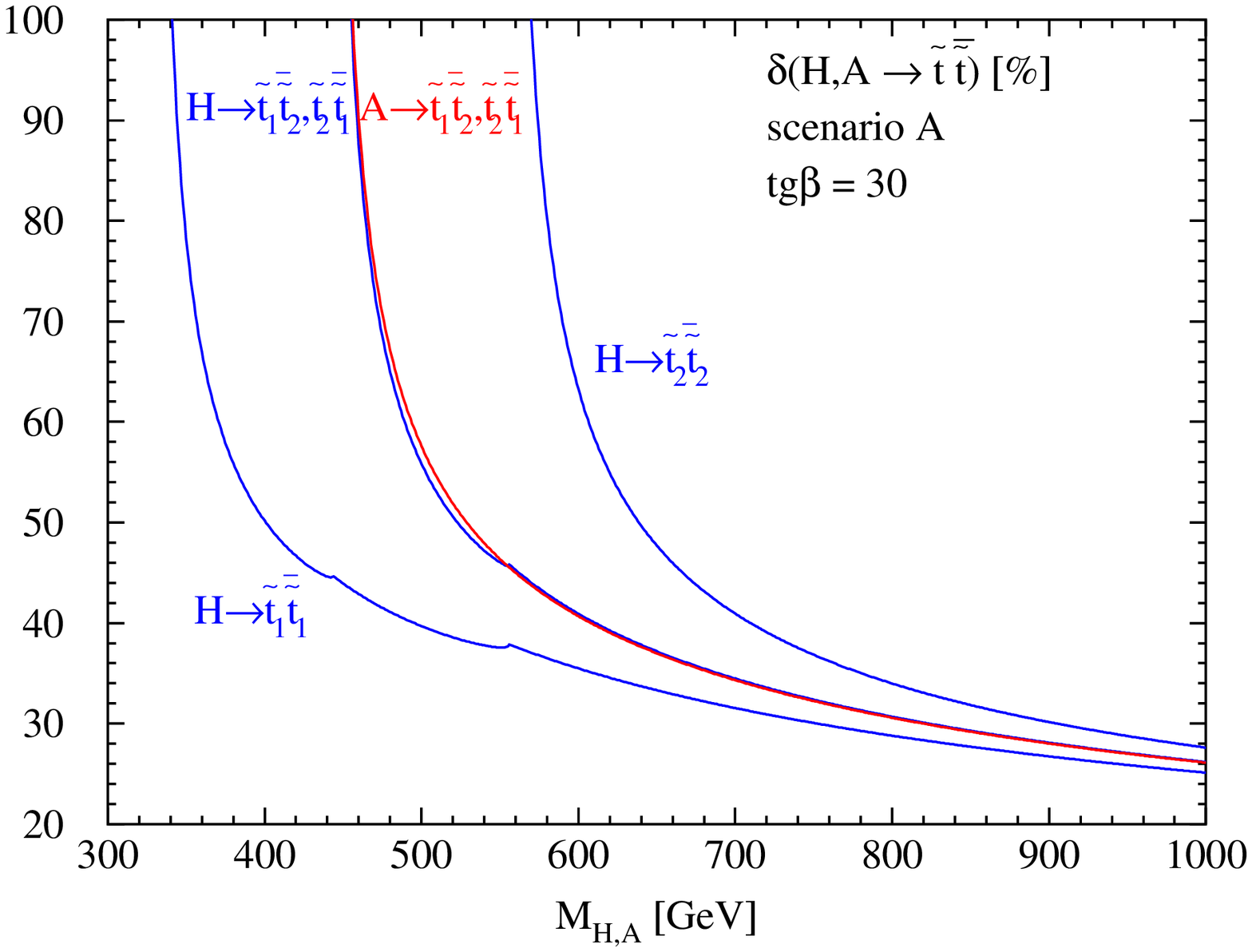}}
\end{picture}
\caption[]{\label{fg:vienna_t} \it SUSY--QCD corrected partial decay
widths (upper) and the relative SUSY--QCD corrections (lower) of the
heavy scalar and the pseudoscalar MSSM Higgs boson decays to stop pairs
as functions of the corresponding Higgs masses for $\tgb=30$ in scenario
A. The full curves include the SUSY--QCD corrections, while the dashed
ones are the LO results. The small kinks originate from the thresholds
of the other stops in the virtual corrections.}
\end{figure}

\begin{figure}[hbtp]
\begin{picture}(100,500)(0,0)
\put(40.0,120.0){\includegraphics{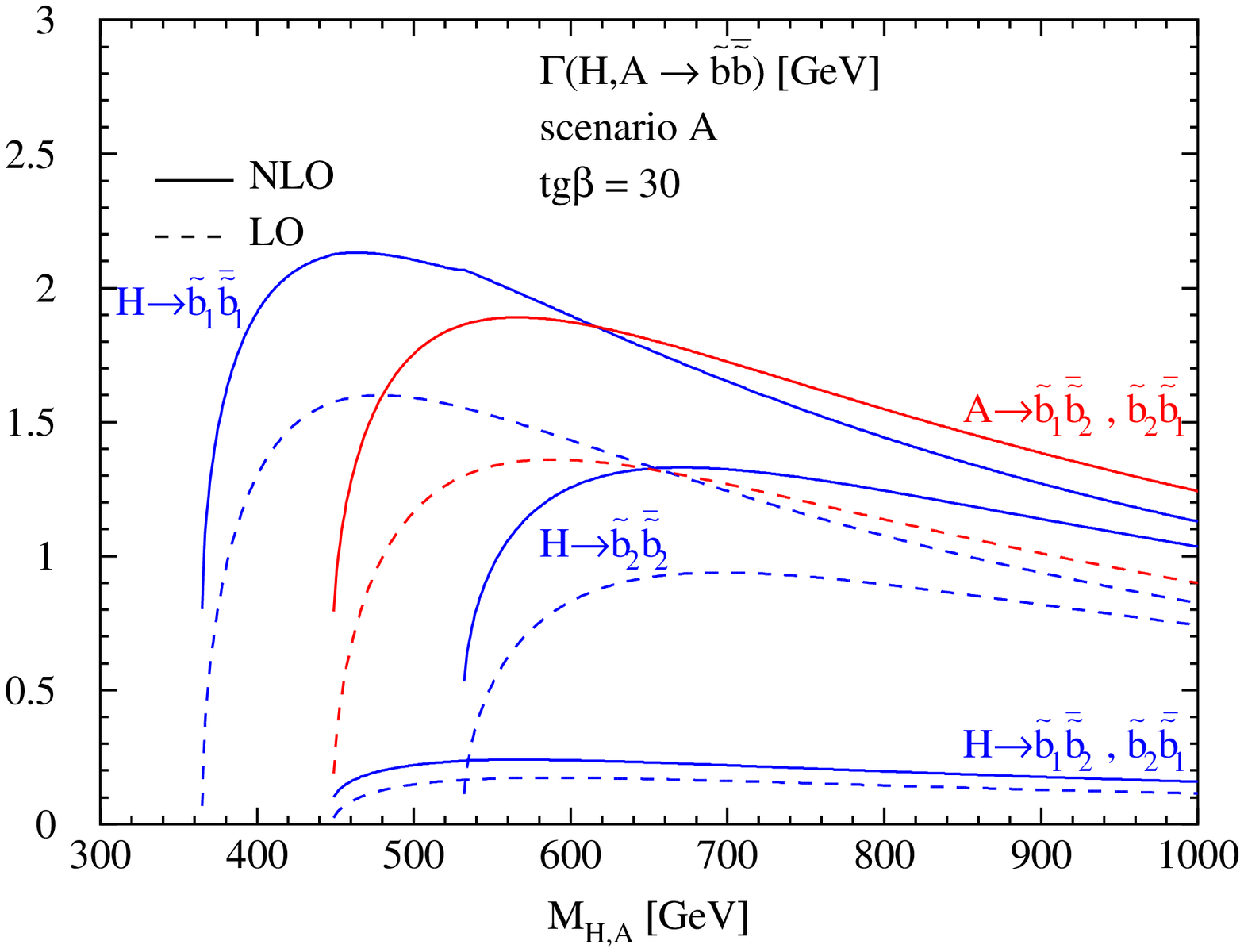}}
\put(40.0,-135.0){\includegraphics{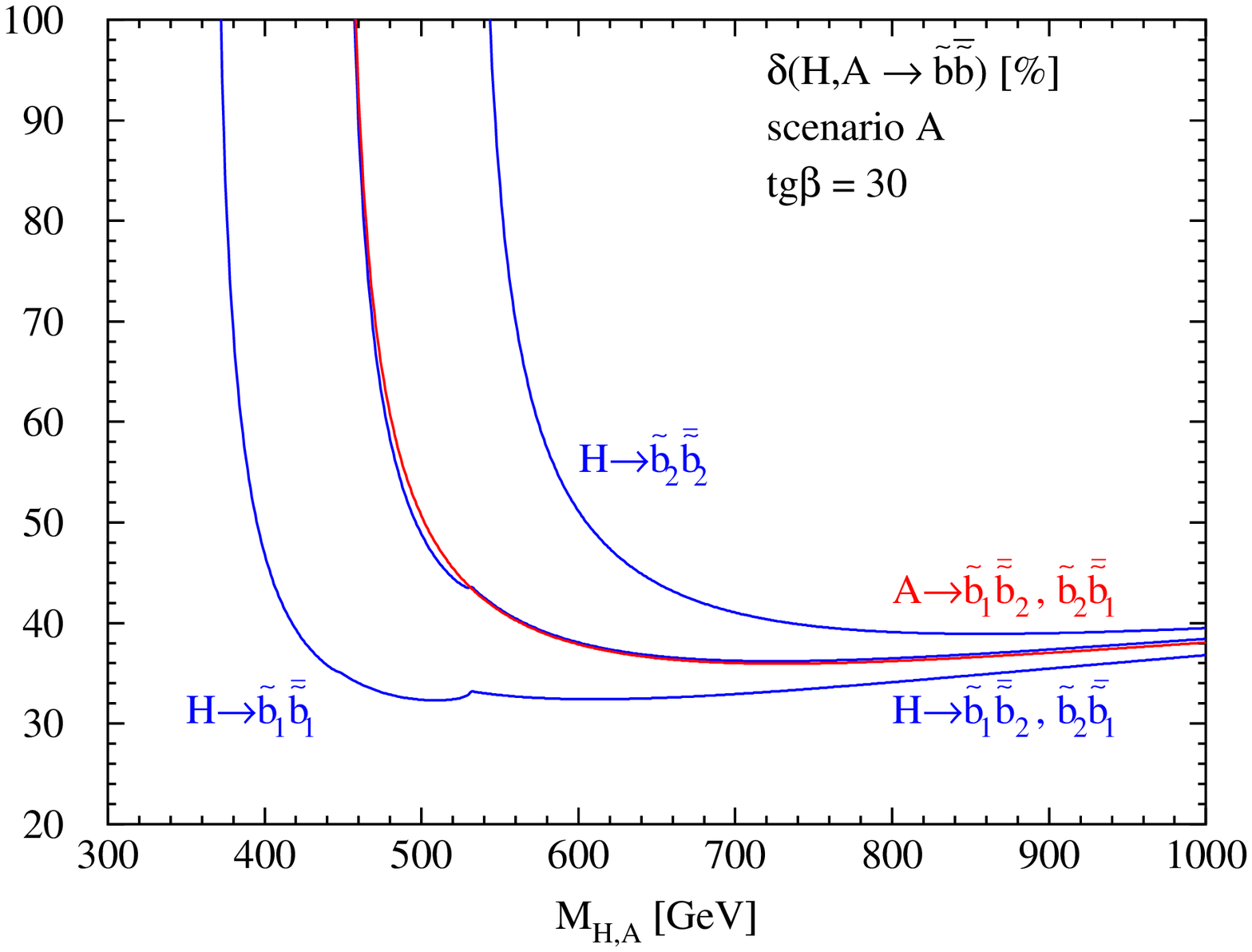}}
\end{picture}
\caption[]{\label{fg:vienna_b} \it SUSY--QCD corrected partial decay
widths (upper) and the relative SUSY--QCD corrections (lower) of the
heavy scalar and the pseudoscalar MSSM Higgs boson decays to sbottom
pairs as functions of the corresponding Higgs masses for $\tgb=30$ in
scenario A. The full curves include the SUSY--QCD corrections, while the
dashed ones are the LO results.  The small kinks originate from the
thresholds of the other sbottoms in the virtual corrections.}
\end{figure}

In Fig.~\ref{fg:vienna_t} we display the results for the partial decay
widths of the heavy neutral Higgs particles into stop pairs and the
relative corrections in scenario A. Since the stops are moderately heavy
all decay modes are kinematically allowed for Higgs masses above about
560 GeV. The partial decay widths range at the few GeV level and turn
out to be similar for all Higgs bosons above the corresponding
thresholds except the scalar Higgs decay widths into non-diagonal stop
pairs which are smaller. Apart from the Coulomb singularities at
threshold the SUSY--QCD corrections are of moderate size in our scheme.
Similar results have been obtained for the neutral Higgs boson decay
widths into sbottom pairs as shown in Fig.~\ref{fg:vienna_b}. The
moderate size of the full SUSY--QCD corrections confirms the proper
absorption of the $\Delta_b$ terms as in Eq.~(\ref{eq:deltab}). If the
trilinear coupling $A_b$ and the bottom quark mass $m_b$ would be
renormalized in the on-shell scheme the SUSY--QCD corrections would
increase the LO results by more than an order of magnitude so that the
result in the on-shell scheme becomes totally unreliable \cite{hsqimp,
walser}. The partial widths of the non-diagonal Higgs decays $H\to
\tilde b_1 \overline{\tilde b}_2, \tilde b_2 \overline{\tilde b}_1$ are
small, since the mixing factor $\cos 2\theta_b$ is small and thus the
contributions of the left-right couplings $g^H_{\tilde b_L \tilde b_R}$
are suppressed, while the diagonal couplings $g^H_{\tilde b_L \tilde
b_L}$ and $g^H_{\tilde b_R \tilde b_R}$ cancel each other to a large
extent in the evaluation of the coupling $g^H_{\tilde b_1 \tilde b_2}$
of Eq.~(\ref{eq:couprot}). Using the scheme of Ref.~\cite{beenakker} for
the mixing angle, i.e.~defining the mixing angle counter terms as
$\delta \theta_q = \Re e\Sigma_{12}(Q^2) /(m_{\tilde q_2}^2
- m_{\tilde q_1}^2)$ with $Q^2 = m_{\tilde q_1}^2$ or $Q^2 = m_{\tilde
  q_2}^2$, the results agree with ours within less than 1\%\footnote{The
results using the tree-level like mixing angle $\tilde \theta_q$ of
Eq.~(\ref{eq:sqmix}) develop sizeable differences to the results in our
scheme for squark masses $m_{\tilde q_{1,2}}$ close to each other, where
the partial decay widths $H\to \tilde q_1 \overline{\tilde q_2}, \tilde
q_2 \overline{\tilde q_1}$ turn out to be negative thus signalizing a
basic problem with the tree-level-like mixing angle.}.

\begin{figure}[hbtp]
\begin{picture}(100,500)(0,0)
\put(40.0,120.0){\includegraphics{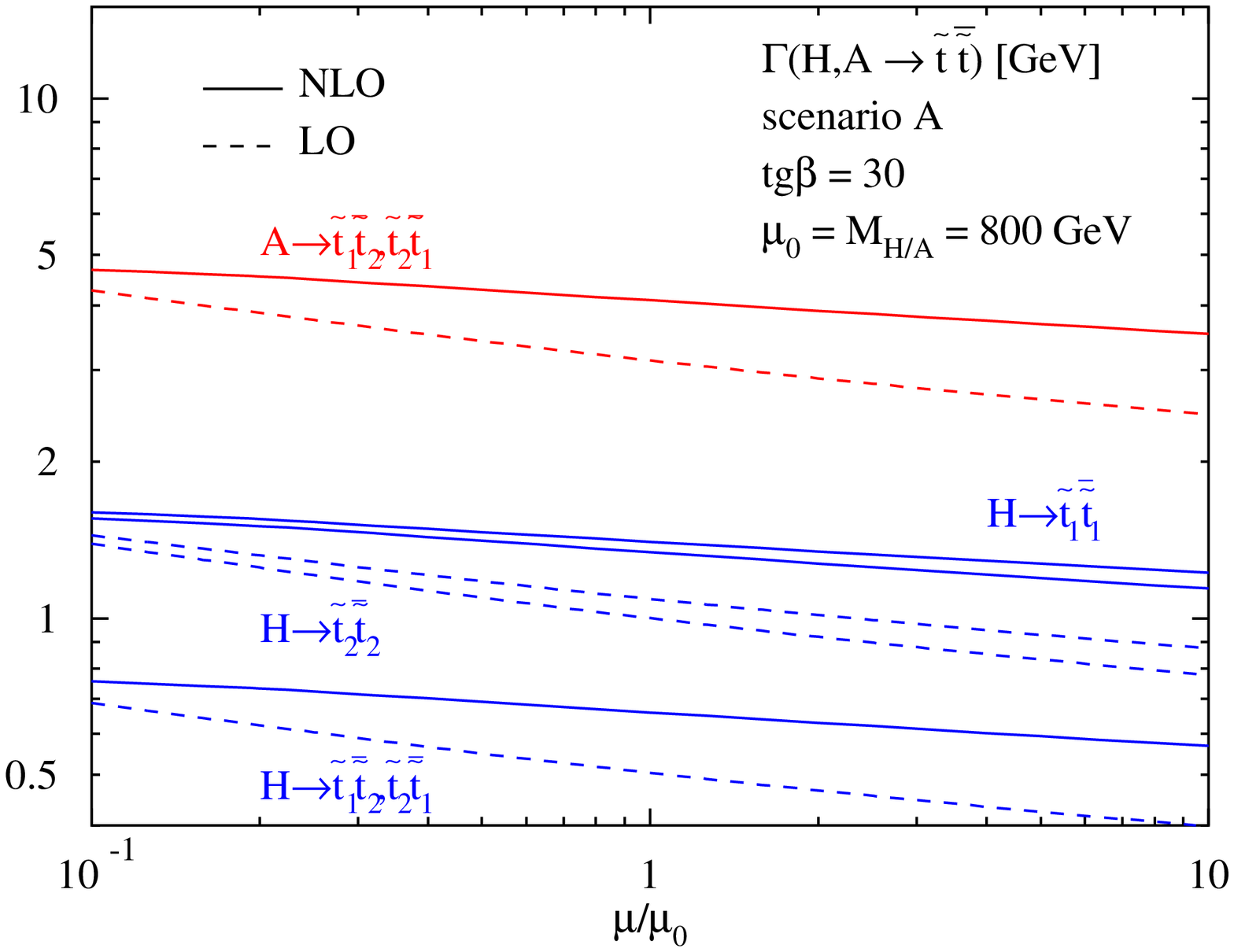}}
\put(40.0,-135.0){\includegraphics{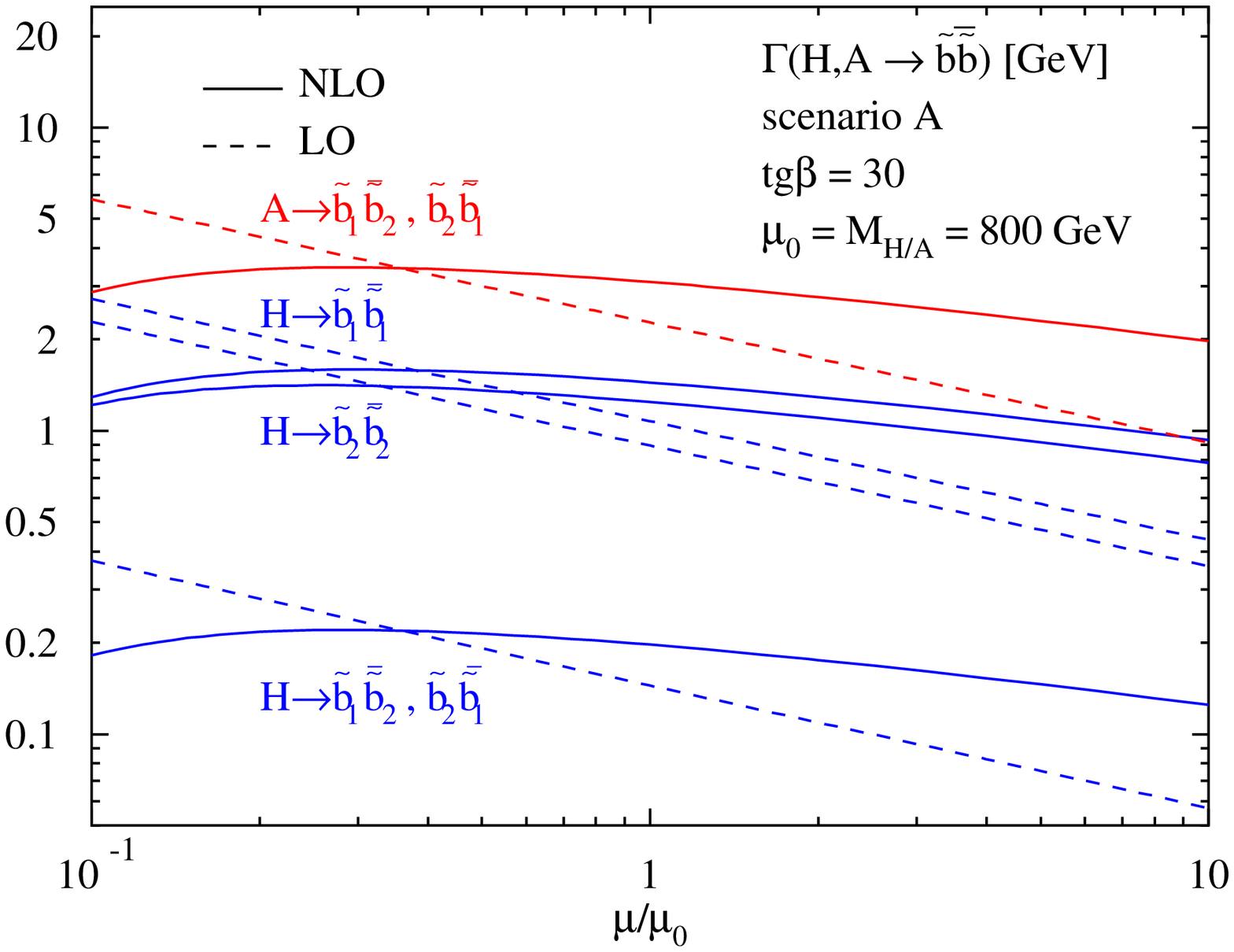}}
\end{picture}
\caption[]{\label{fg:scale_v} \it Scale dependences of the partial decay
widths into stop pairs (upper) and sbottom pairs (lower) as functions of
the renormalization scale in units of the central scale $\mu_0=M_{H/A}$
in scenario A. The broken lines show the leading-order scale
dependences and the full curves the NLO results.}
\end{figure}
In order to obtain an estimate of the residual theoretical uncertainties
we show the renormalization scale dependence of the partial decay widths
into stop and sbottom pairs in Fig.~\ref{fg:scale_v}.  The scale
dependences of the partial decay widths into stop and sbottom pairs are
significantly reduced at NLO. Varying the renormalization scale by a
factor of 2 around the central scale $\mu_R=M_{H/A}$ the theoretical
uncertainties can be estimated as 5--10\% for decays into stop and
sbottom pairs at NLO.

\begin{figure}[hbtp]
\begin{picture}(100,500)(0,0)
\put(40.0,120.0){\includegraphics{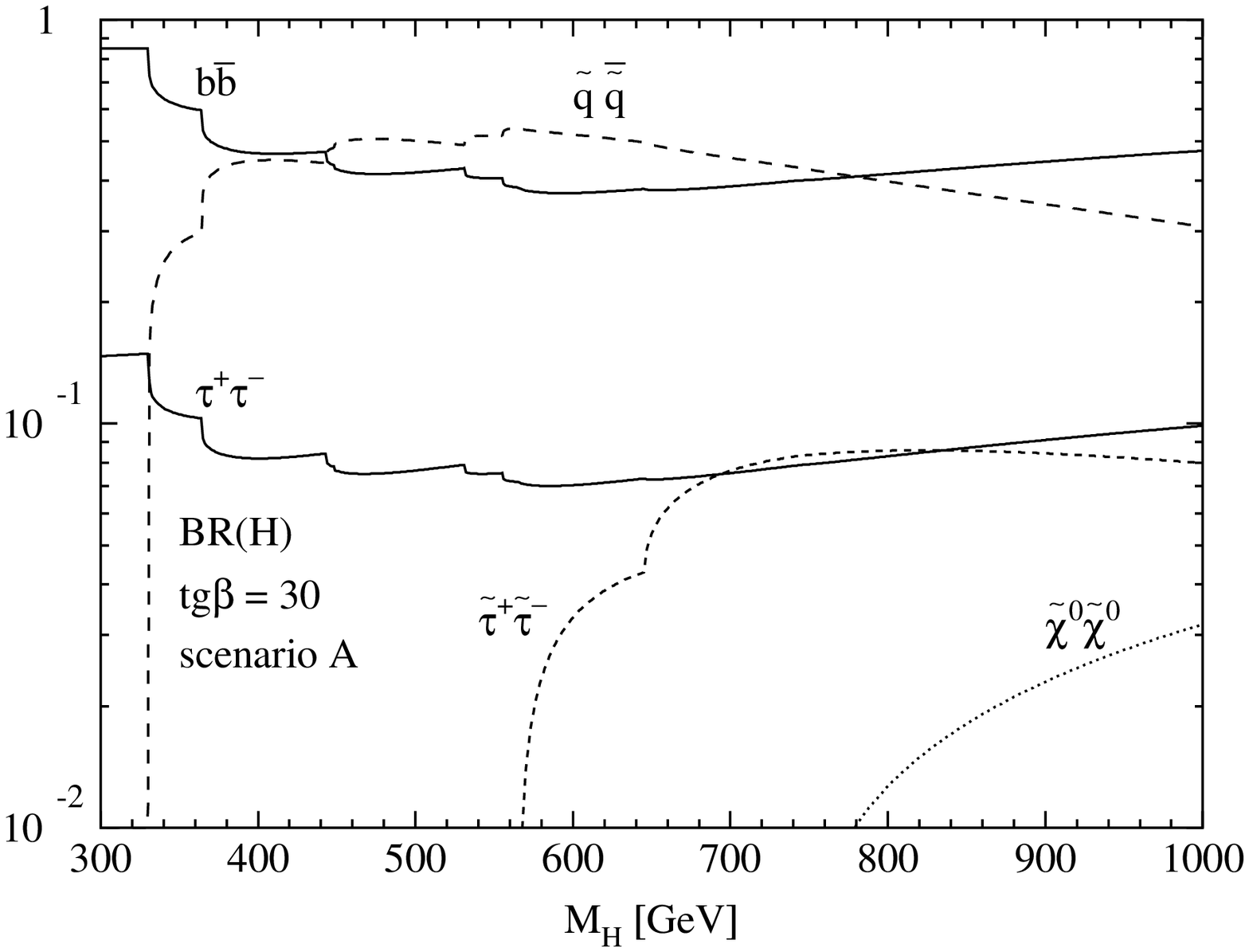}}
\put(40.0,-135.0){\includegraphics{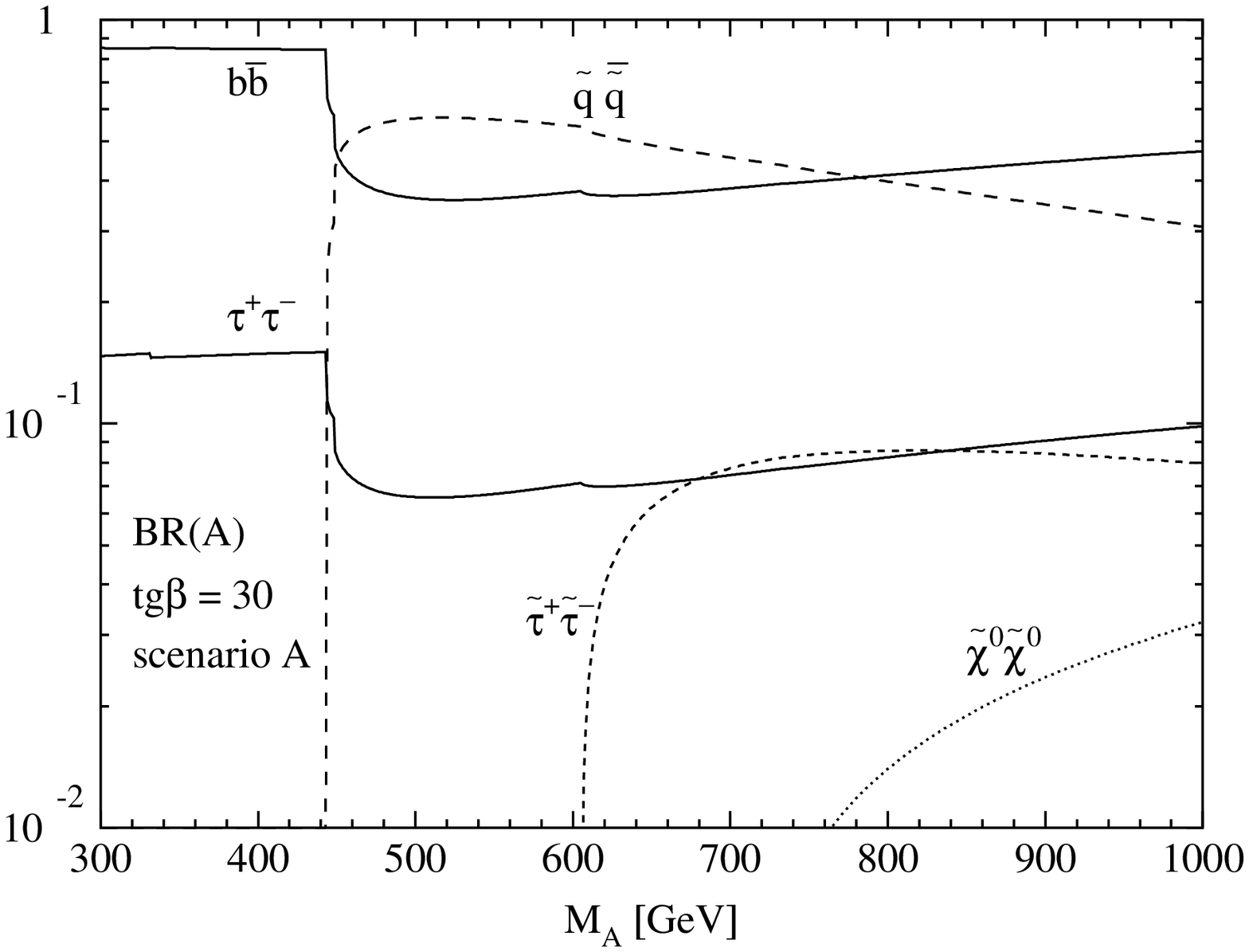}}
\end{picture}
\caption[]{\label{fg:br_v} \it Branching ratios of the heavy scalar $H$
and the pseudoscalar $A$ MSSM Higgs bosons in scenario A as functions of
the corresponding Higgs masses. The curves for decays into neutralinos
$\tilde \chi^0 \tilde \chi^0$ and squarks $\tilde q \overline{\tilde q}$
represent the corresponding sums over all possible mass eigenstates.}
\end{figure}

Fig.~\ref{fg:br_v} displays the corresponding branching ratios of the
dominant decay modes of the heavy scalar and the pseudoscalar Higgs
boson in scenario A. It is clearly visible that the decays into stop and
sbottom pairs belong to the dominant decay modes for masses above about
300--400 GeV, i.e.~where they are kinematically allowed, reaching branching
ratios of up to about 50\% in total. Moreover, in scenario A the decay
modes into $\tilde \tau$'s and neutralinos play a significant role, too.

\begin{figure}[hbtp]
\begin{picture}(100,500)(0,0)
\put(40.0,120.0){\includegraphics{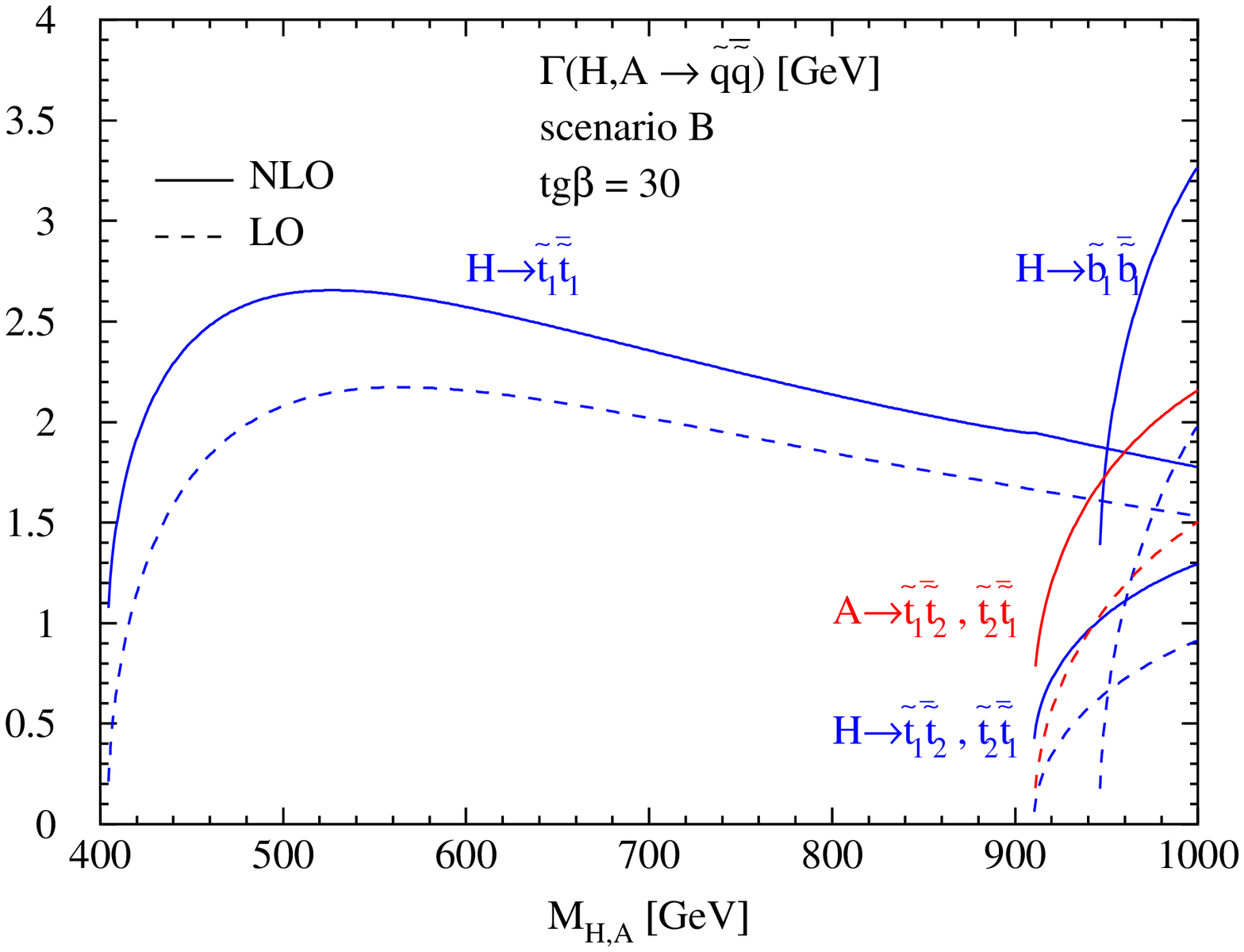}}
\put(40.0,-135.0){\includegraphics{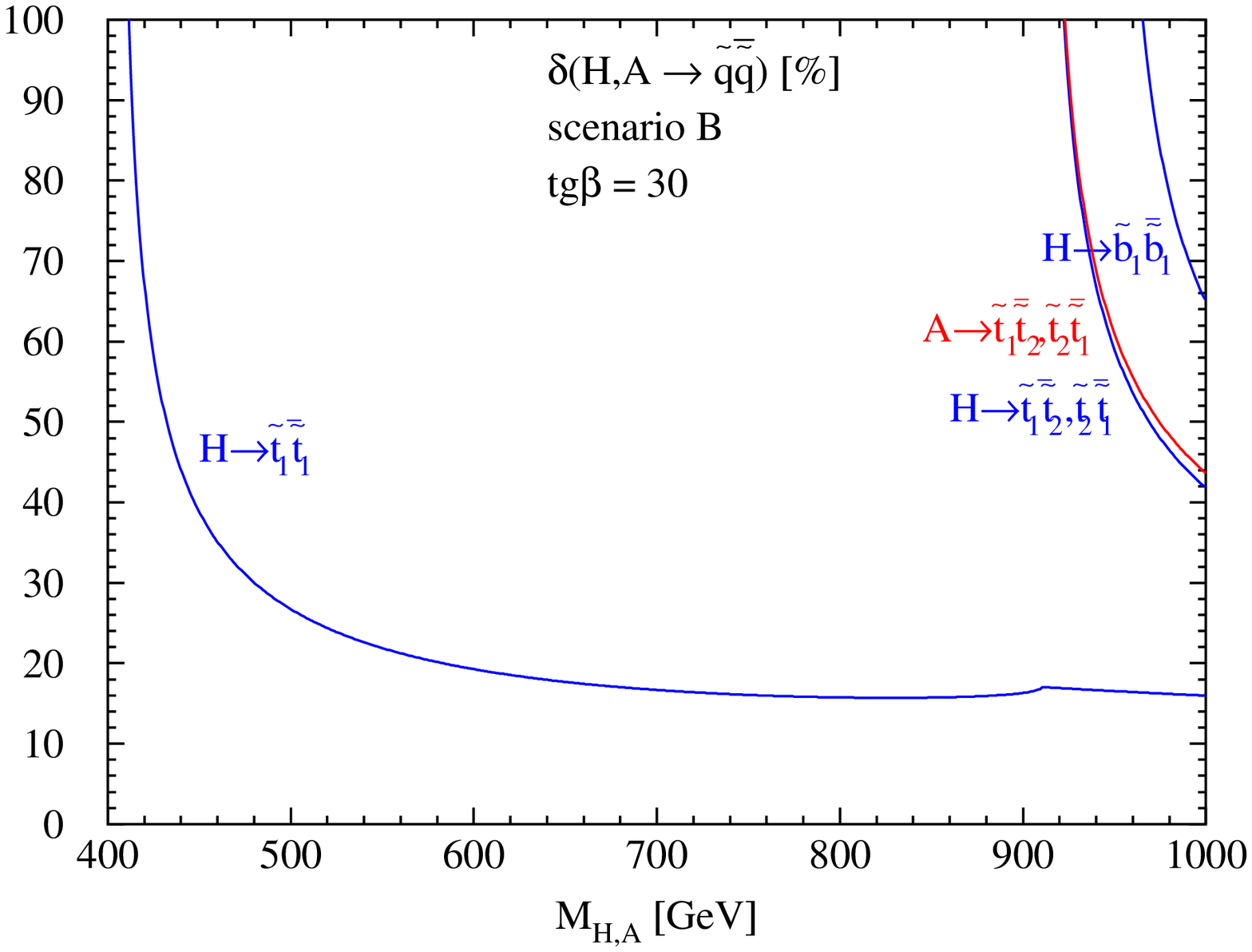}}
\end{picture}
\caption[]{\label{fg:scen_b} \it SUSY--QCD corrected partial decay
widths (upper) and the relative SUSY--QCD corrections (lower) of the
heavy scalar and the pseudoscalar MSSM Higgs boson decays to stop and
sbottom pairs as functions of the corresponding Higgs masses for
$\tgb=30$ in scenario B. The full curves include the SUSY--QCD
corrections, while the dashed ones are the LO results. The small kinks
originate from the thresholds of the other stops in the virtual
corrections.}
\end{figure}

In Fig.~\ref{fg:scen_b} we display the final results for the scenario B.
Since the masses $m_{\tilde t_2}, m_{\tilde b_{1,2}}$ are close to or
larger than 500 GeV, the decay modes $H\to  \tilde t_2 \overline{\tilde
t}_2, \tilde b_1 \overline{\tilde b}_2, \tilde b_2 \overline{\tilde
b}_1, \tilde b_2 \overline{\tilde b}_2$ are kinematically forbidden for
$M_H\le 1$ TeV.  The partial decay widths of decays into stop and
sbottom pairs range at the few GeV level as in scenario A once they are
kinematically allowed. It can clearly be inferred from
Fig.~\ref{fg:scen_b} that the SUSY--QCD corrections are of moderate size
apart from the threshold regions where the Coulomb singularities enhance
the size of the corrections. It should be noted that the results with
the sbottom mixing angle renormalized via the tree-level relation of
Eq.~(\ref{eq:sqmixct1}) would lead to unphysical negative partial decay
widths for heavy scalar Higgs decays $H\to \tilde b_1 \overline{\tilde
b}_2, \tilde b_2 \overline{\tilde b}_1$ for Higgs masses larger than 1
TeV, where these decay channels open up, so that this scheme is strongly
disfavoured.  These negative contributions can be traced back to the
uncancelled artificial singularity for $m_{\tilde b_1}\sim m_{\tilde
b_2}$ in the finite remainders $\Delta \theta_b$ of
Eq.~(\ref{eq:deltath}), if $\delta\theta_b$ is replaced by the counter
term $\delta\tilde\theta_b$ of Eq.~(\ref{eq:sqmixct1}) \cite{hsqimp}.

\begin{figure}[hbtp]
\begin{picture}(100,500)(0,0)
\put(40.0,120.0){\includegraphics{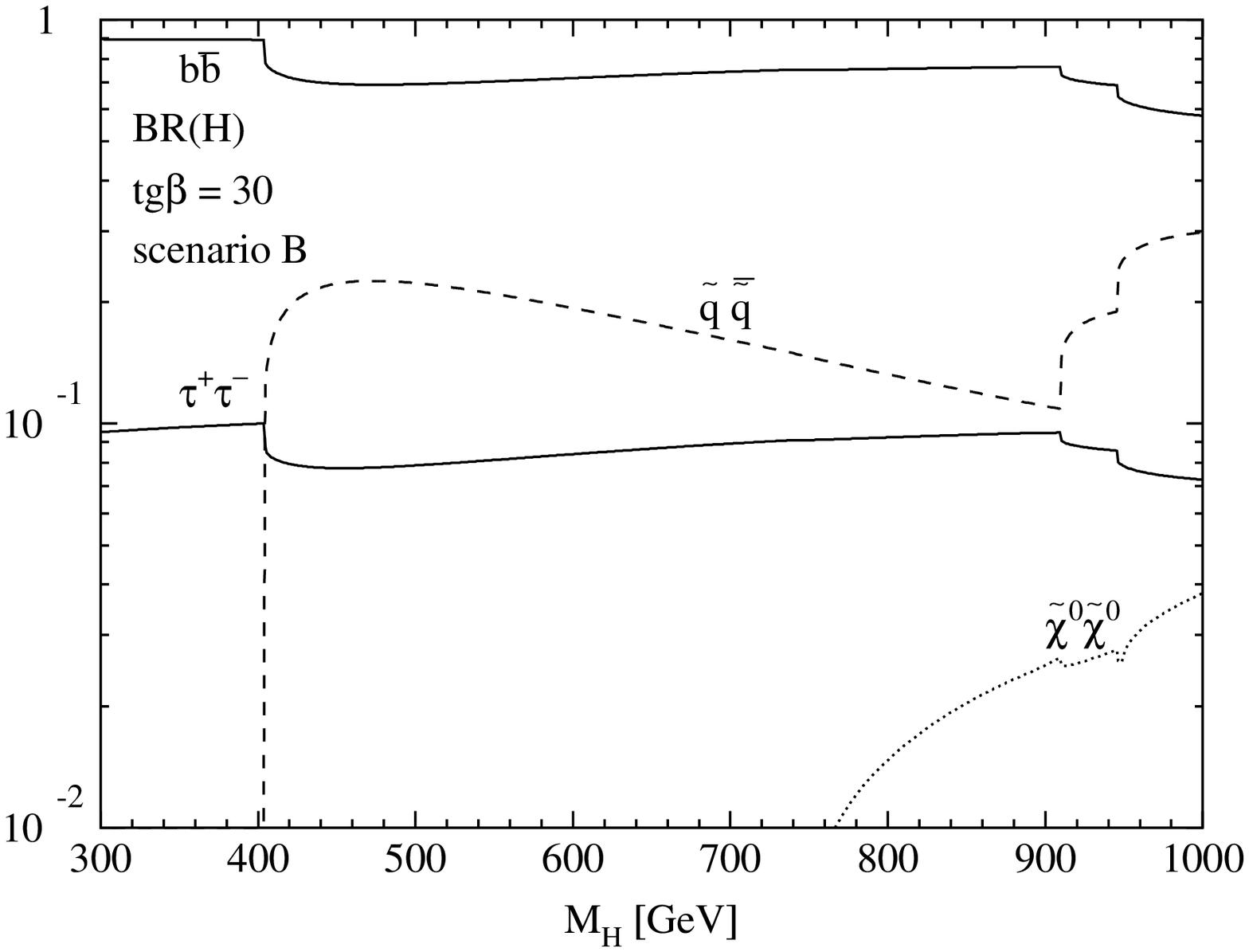}}
\put(40.0,-135.0){\includegraphics{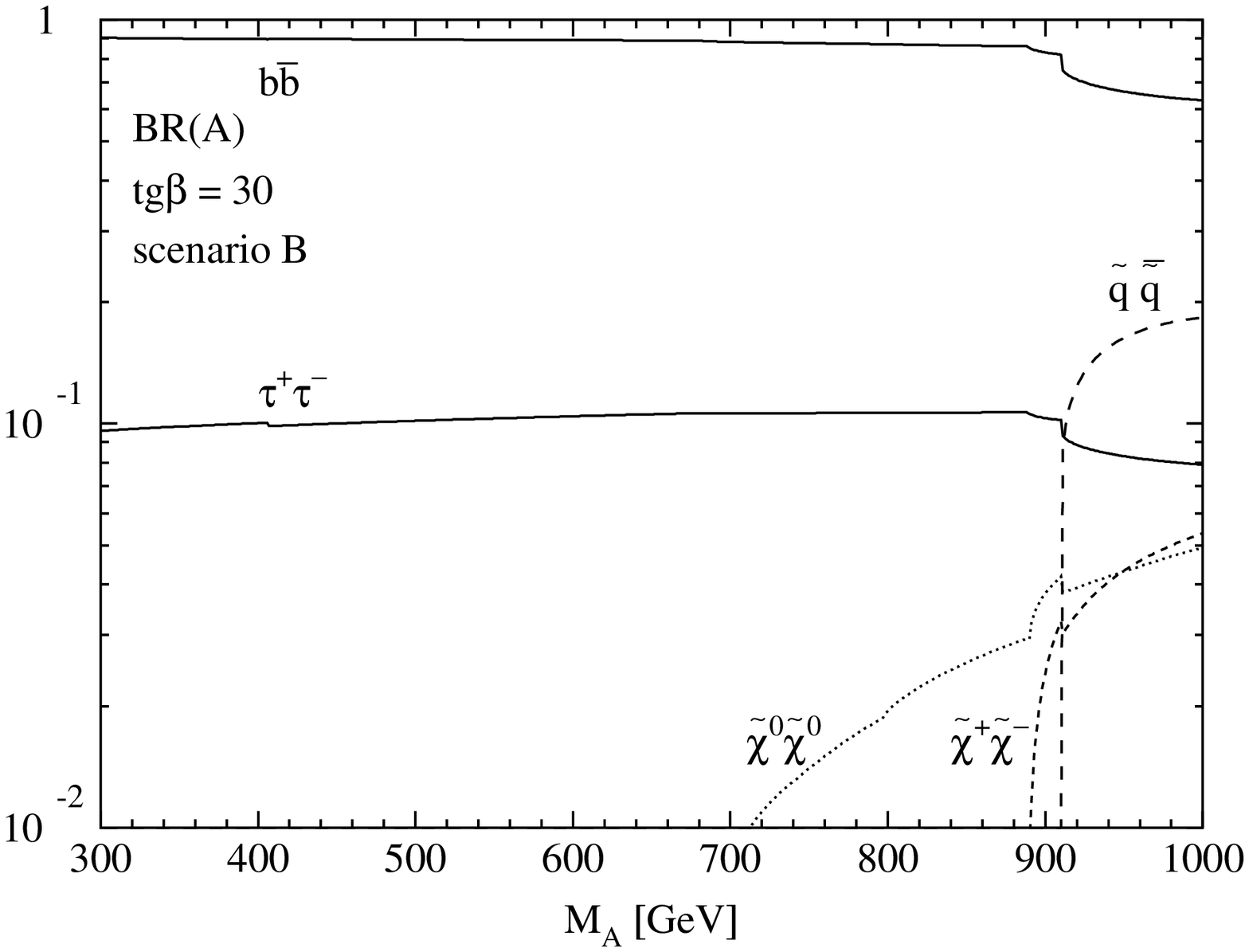}}
\end{picture}
\caption[]{\label{fg:br_b} \it Branching ratios of the heavy scalar
$H$ and the pseudoscalar $A$ MSSM Higgs bosons in scenario B as
functions of the corresponding Higgs masses. The curves for decays into
charginos $\tilde \chi^+ \tilde \chi^-$, neutralinos $\tilde \chi^0
\tilde \chi^0$ and squarks $\tilde q \overline{\tilde q}$ represent the
corresponding sums over all possible mass eigenstates.}
\end{figure}

In Fig.~\ref{fg:br_b} we show the branching ratios of the heavy scalar
and the pseudoscalar Higgs boson in scenario B as functions of the
corresponding Higgs masses. As in scenario A the decay modes into stop
and sbottom pairs play a significant role, once they are kinematically
allowed. This starts to be the case for the heavy scalar Higgs boson $H$
already for masses above about 400 GeV, where the decay into light stop
mass eigenstates opens up. The pseudoscalar Higgs boson can only decay
into mixed pairs of a light and heavy stop and sbottom eigenstates so
that these decay modes open up for masses above 900 GeV. The decays into
charginos and neutralinos contribute significantly for larger Higgs
masses.

For smaller values of $\tgb$ the decay modes $H\to hh,t\bar t$ and $A\to
Zh, t\bar t$ play a significant role, while the decays into sbottom
pairs are usually suppressed. However, the neutral Higgs boson decays
into stop pairs still play a significant role and can even be the
dominant heavy Higgs boson decays in a large Higgs mass range. The
SUSY--QCD corrections to these decay modes are of similar size as in the
scenarios with large values of $\tgb$ analyzed in this work.

\section{Conclusions} \label{sc:conclusions}
In this work we have analyzed the NLO SUSY--QCD corrections to heavy
MSSM Higgs decays into stop and sbottom pairs and elaborated a
consistent scheme and setup for the theoretical NLO results and the
corresponding input parameters, i.e.~the squark masses and their
couplings to the Higgs bosons. A reliable determination of these
corrections requires a proper treatment of contributions which are
enhanced by $\tgb$. The derivation of the input squark masses and
couplings with NLO accuracy requires a modification of the tree-level
relations among the stop and sbottom masses by higher-order corrections,
if they should correspond to the physical pole masses. Since the size of
these modifications depends on the MSSM parameters and pole masses
themselves an iterative procedure should be applied. These iterations,
however, do not converge for all scenarios so that we restricted the
evaluation to a single iteration step. The final setup of the input
parameters matches NLO accuracy consistently. A preferred choice of
the scheme for the stop and sbottom mixing angles is the on-shell
definition via the anti-Hermitian counter term which removes spurious
singularities for nearly mass degenerate cases which drive the
off-diagonal decays $H\to \tilde b_1 \overline{\tilde b}_2$ to negative
and thus unphysical values. The final NLO corrections are of moderate
size apart from the expected Coulomb singularities at the decay
thresholds. The residual theoretical uncertainties at NLO have been
estimated to be 5--10\%. \\

\noindent
{\bf Note added in proof.}
During completion of this work Ref.~\cite{heidi} appeared which
discusses several options for the renormalization of the stop and
sbottom sector with NLO accuracy. Their treatment of the stop and
sbottom sectors is different from our approach in particular for the
effective bottom mass and the squark mixing angles. \\

\noindent
{\bf Acknowledgments.} \\
We are grateful to A.~Djouadi, M.~M\"uhlleitner, H.~Rzehak and
P.M.~Zerwas for carefully reading our manuscript and useful comments.

\end{document}